\documentclass[11pt,letterpaper]{article}
\usepackage[T1]{fontenc}
\usepackage{lmodern}
\usepackage[margin=1in]{geometry}
\usepackage{amsmath,amssymb,amsthm,mathtools}
\usepackage{booktabs,longtable,array,tabularx,enumitem}
\usepackage{microtype}
\usepackage{graphicx}
\usepackage{placeins}
\usepackage[numbers]{natbib}
\usepackage[hidelinks]{hyperref}
\usepackage[nameinlink,noabbrev,capitalise]{cleveref}
\usepackage{xcolor}

\newcommand{\N}{\mathbb N}
\newcommand{\Z}{\mathbb Z}
\newcommand{\Q}{\mathbb Q}
\newcommand{\PP}{\mathbb P}
\newcommand{\vp}{v_p}
\DeclareMathOperator{\Mat}{Mat}
\DeclareMathOperator{\GL}{GL}
\DeclareMathOperator{\im}{im}
\newtheorem{theorem}{Theorem}[section]
\newtheorem{lemma}[theorem]{Lemma}
\newtheorem{proposition}[theorem]{Proposition}
\newtheorem{corollary}[theorem]{Corollary}
\theoremstyle{definition}
\newtheorem{definition}[theorem]{Definition}
\theoremstyle{remark}
\newtheorem{remark}[theorem]{Remark}
\numberwithin{equation}{section}
\setlist{itemsep=2pt,topsep=4pt}
\title{Mortality Is Undecidable for $2\times2$ Integer Matrices}
\author{Thanin Quartz$^{\dagger}$ \quad Ivan O. Shevchenko$^{\ddagger}$ \quad Hans De Sterck$^{\S}$ \quad Jun Liu$^{\P}$\\[0.5em]
{\normalsize University of Waterloo, Department of Applied Mathematics}}
\date{October 2026}
\begin{document}
\pagenumbering{roman}
\maketitle
\begin{abstract}
The matrix mortality problem asks whether a given finite family of
matrices can be multiplied, in any order and with possible repetition, to get the zero matrix. It is one of the most basic decision problems about matrix products, and it arises naturally in the study of switched linear systems and of finite automata. Mortality is trivially decidable in dimension one and has been known to be undecidable for integer matrices in dimensions three and higher since 1970; only the problem in dimension $2$ is left open. We resolve this problem by proving that
mortality is undecidable for families of 5,799 distinct $2\times2$
integer matrices. All but one of the matrices are fixed and nonsingular, with determinants $\pm p$ for a single fixed prime $p$. The input affects only the remaining rank-one matrix, which encodes this input into its kernel.

Our proof encodes the computation of a fixed universal machine into products of $2\times2$ matrices. We represent each configuration of the computation by a single rational number, which we identify with a line through the origin in $\Q^2$, and each nonsingular matrix carries out a small piece of one computation step. The main difficulty is that the matrices may be multiplied in any order, so we design them to check their own work: a ``wrong'' step puts a factor of a prime number $p$ into the denominator of the rational number corresponding to the current computation step, where no later matrix can remove it. As a result, products of these matrices can go from one encoded configuration to another only by following a valid computation. The rank-one matrix then turns the question of whether the machine halts into the question of whether some product equals
zero. A companion formalization in the Lean proof assistant verifies
the complete reduction.
\end{abstract}

\vfill
\noindent\begin{minipage}{\linewidth}
\footnotesize\raggedright
$^{\dagger}$\,Email: \href{mailto:tquartz@uwaterloo.ca}{tquartz@uwaterloo.ca}\\
$^{\ddagger}$\,Email: \href{mailto:i3shevch@uwaterloo.ca}{i3shevch@uwaterloo.ca}\\
$^{\S}$\,Email: \href{mailto:hans.desterck@uwaterloo.ca}{hans.desterck@uwaterloo.ca}\\
$^{\P}$\,Email: \href{mailto:j.liu@uwaterloo.ca}{j.liu@uwaterloo.ca}
\end{minipage}

\clearpage
\tableofcontents
\clearpage
\pagenumbering{arabic}
\addcontentsline{toc}{part}{Part I. Main results and technical overview}
\begin{center}\large\bfseries Part I. Main results and technical overview\end{center}
\vspace{-0.5\baselineskip}
\section{Introduction and main result}
\label{conf:introduction}\label{sec:introduction}

Matrix mortality is a decision problem that asks, given a finite set of $n\times n$ matrices with integer entries, whether the zero matrix can be expressed as a nonempty product of matrices from this set. If so, the family is called \emph{mortal}. The problem has its roots in computability theory and arises in control theory, where it is the same as asking whether a linear switched system can be ``uniformly'' driven to zero \citep{EgerstedtBlondel2002}, and in automata theory, where for nonnegative transition matrices it is the same as asking whether some word has no run from any state \citep[Section~2.1]{KieferRyzhikov2025}.

In dimension one, mortality is decidable by checking whether the family contains zero, and in dimensions three and higher it has been known to be undecidable since 1970 \citep{Paterson1970}. The two-dimensional case has remained open \citep[Section~1]{BournezBranicky1999}, \citep[Open Problem~1]{BournezBranicky2002}. Partial results establish decidability for families of two rational matrices \citep[Theorem~4]{BournezBranicky2002}, for finite rational families containing at most one nonsingular matrix \citep[Theorem~1]{Heckman2019}, and for integer families in which every matrix has determinant in $\{0,1,-1\}$ \citep{NuccioRodaro2008}. Appendix~\ref{app:scope} surveys further results.

A major difficulty in proving undecidability for matrix mortality is figuring out how to encode computation with matrix multiplications. Since given matrices can be multiplied in any order and used arbitrarily often, when they represent transitions of a machine there is nothing which forces a product to follow a legal execution; yet, we somehow need to distinguish between zero and non-zero products in relation to whether the machine halts or not. For dimension two, we have the additional difficulty that there are no spare coordinates for bookkeeping. This is related to the fact that one-dimensional subspaces of $\Q^2$ are determined by a single rational number, and so corresponding machine configurations must be packed into one number. Yet another difficulty is imposed on us by the fact that if a collection of nonsingular matrices all have integer entries and determinant one, then there is an algorithm to decide whether a product of their fractional linear transformations maps one given rational point to another \citep[Theorem~14]{PotapovSemukhin2019}. As such, we must be careful to ensure that not all of the matrices in our construction have determinants equal to one.

We resolve the two-dimensional case by proving that mortality is
undecidable for $2\times2$ integer matrices, in the following strong
form. Write $\Mat_2(K)$ to denote the $2\times2$ matrices with entries in
$K$, and $\im R$ and $\ker R$ for the column space and the kernel of
$R$.

\begin{theorem}[Mortality with one variable matrix]
\label{conf:main}\label{thm:main}
There exist a prime $p$, a fixed finite family
$\mathcal G\subseteq\Mat_2(\Z)$ of nonsingular matrices with
determinants in $\{p,-p\}$, and a fixed rational line
$L\subseteq\Q^2$ such that no algorithm decides, given a nonzero
rank-one matrix $R\in\Mat_2(\Z)$ with $\im R=L$, whether
$\mathcal G\cup\{R\}$ is mortal.
\end{theorem}

\begin{corollary}\label{conf:count}
The family $\mathcal G$ in \cref{conf:main} can be chosen to have
exactly 5,798 distinct matrices. Consequently, mortality is
undecidable for families of exactly 5,799 two-by-two integer matrices.
\end{corollary}

All but one of these matrices are fixed in advance, and the input enters only through the ``kernel line'' of $R$, whose image is the fixed line $L$. The hardness of mortality in dimension two thus resides in the choice of one rational line. In contrast, the membership of a prescribed nonsingular matrix in the products of nonsingular integer matrices is decidable \citep[Theorem~2.1]{PotapovSemukhin2017Nonsingular}. In Appendix~\ref{app:scope} we explain why that algorithm does not apply to a prescribed line. In our construction we represent the configurations of a fixed program by rational lines, and build ``checks'' for this program's control states and arithmetic updates into certain matrix actions. These checks ensure that a product of the nonsingular matrices can map one encoded configuration to another only if the program can reach the second configuration from the first. The arbitrary-looking number 5,799 in \cref{conf:count} is the size of the family of matrices our construction produces from one particular universal machine, and it is not optimized. Section~\ref{conf:overview} gives an overview of the construction. Part~II contains the detailed proofs and generator tables, followed by a description of a companion formal verification.

\section{Overview of the proof}
\label{conf:overview}

\begin{figure}[!htbp]
  \centering
  \includegraphics[width=\linewidth]{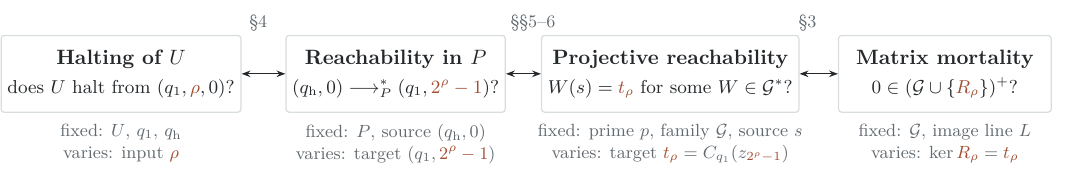}
  \caption{The three equivalences proving \cref{conf:main}, established by \cref{conf:program}, \cref{conf:compiler} with \cref{conf:instruction-implementations}, and \cref{conf:rank-one}, respectively. Beneath each step of the proof, we list the fixed objects and the object that varies with the input $\rho$.}
  \label{fig:construction-overview}
\end{figure}

To prove \cref{conf:main}, we start from the undecidable halting problem for a fixed two-counter machine $U$ (\cref{prog:source}). Figure~\ref{fig:construction-overview} shows the three equivalences that connect this problem to matrix mortality. We first express this halting problem as reachability in a fixed arithmetic program $P$, then implement $P$ with a fixed matrix family $\mathcal G$. The reduction is completed by appending a rank-one matrix $R_\rho$ to the family $\mathcal G$. The figure also keeps track of what remains fixed and what depends on the input $\rho$.

\paragraph{From the machine to a one-counter program (Section~\ref{conf:program-section}).}
We first encode $U$'s two counters in a single integer $n$ using prime factorization, which will later be represented by a rational number. The two counter values are the exponents of $2$ and $3$ in the prime factorization of the number $n+1$, and we translate their increments, decrements, and zero tests into arithmetic updates and tests on $n$. The simulation starts from $(q_1,2^\rho-1)$, representing $U$'s initial state and counter values $(\rho,0)$. However, our reduction requires a fixed starting configuration corresponding to the fixed image of the matrix $R$ in \cref{conf:main}. Running the simulation backwards (i.e., reversing the transitions) would move the input into the target, but the counter may have different values when $U$ halts on different inputs. We therefore add a routine at the halt state $q_{\mathrm h}$ that reduces the counter to zero, allowing every halting computation to reach the same endpoint $(q_{\mathrm h},0)$. Reversing the resulting transitions gives the one-counter program $P$, with this common endpoint as its fixed ``source'' and the simulation's original starting configuration as its ``target''. This target is reachable exactly when $U$ halts on input $\rho$ (\cref{conf:program}).

\paragraph{From the program to projective reachability
(Sections~\ref{conf:control} and~\ref{conf:arithmetic}).}
Building on the one-counter program $P$ from Section~\ref{conf:program-section}, we construct matrices that simulate its instructions. We represent each configuration of the program by a rational number corresponding to a rational line, on which the matrices act. For a prime number $p$, we store the counter value $n$ in a rational number
$z_n$ whose base-$p$ digits, starting from the lowest, repeat the
block $0^n1$; the number of zeros records $n$. A $2\times2$ matrix can
delete the lowest digit of a number, and each such matrix is built to
expect a particular digit: if the actual digit is different, a factor
of $p$ appears in the denominator and marks the error. Deleting the
zeros of one block and then its final $1$ returns to $z_n$, so the
matrices can read and test the counter without changing it. To record
the state as well, we shift the number by an integer that depends on
the state $q$, and call the result the configuration code $C_q(z_n)$.
The shifts are chosen so that the residue of the code modulo $p$
reveals the state; as a result, a matrix intended for one state
rejects a configuration in any other state. One instruction may require several matrix multiplications, so we implement it by a finite graph whose vertices record its intermediate stages. Each edge supplies a single matrix that removes the shift for its source state, performs the prescribed arithmetic operations and digit deletion, and adds the shift for its destination state, as in \eqref{conf:generator}. Loops in the graph let the number of multiplications depend on $n$, using the same fixed matrices. A path from the source of $P$ to the target then gives a matrix product mapping the encoded source $s=C_{q_{\mathrm h}}(z_0)$ to the encoded target $t_\rho=C_{q_1}(z_{2^\rho-1})$.

\paragraph{From projective reachability to mortality
(Section~\ref{conf:reduction}).}
The matrices in $\mathcal G$ are nonsingular, so their products cannot be zero. We therefore add a rank-one integer matrix $R_\rho$, choosing its image to be the fixed rational line $L$ represented by a rational number $s$ and its kernel to be the line represented by a rational number $t_\rho$. For a product $W$ of matrices from $\mathcal G$, suppose $W$ sends the line $L$ onto $\ker R_\rho$. Then since $R_\rho$ maps the plane onto $L$ and also $R_\rho$ maps $\ker R_\rho$ to zero, we have $R_\rho WR_\rho=0$. Conversely, \cref{conf:rank-one} shows that any zero product of matrices from $\mathcal G\cup\{R_\rho\}$ yields a product $W$ of matrices from $\mathcal G$ satisfying $W(s)=t_\rho$ (here, we are identifying $s$ and $t_{\rho}$ with the lines they are representing). Thus $\mathcal G\cup\{R_\rho\}$ is mortal exactly when $U$ halts on input $\rho$. This keeps $\mathcal G$ and $L$ fixed and places the input entirely in the kernel of $R_\rho$, as required by \cref{conf:main}.

\vspace{12pt}

The most delicate step in this whole scheme is the converse of the second equivalence in \cref{fig:construction-overview}: a product of matrices from $\mathcal G$ must be able to ``connect'' two encoded configurations only by following a \textit{valid} computation of $P$. Since every matrix in $\mathcal G$ can be used at any step, the instruction graphs alone do not restrict a product. We need to construct the matrices so that a wrong state or digit choice prevents any further product from $\mathcal G$ from reaching an encoded target (\cref{conf:error,conf:guard}). A product may still pass these checks while repeating an operation the wrong number of times. Section~\ref{conf:arithmetic} checks whether these repetitions give the required arithmetic update and ensures that a product cannot switch instructions to avoid the tests. Thus every product from $\mathcal G$ connecting encoded configurations corresponds to a valid computation of $P$ (\cref{conf:compiler,conf:instruction-implementations}).
\section{Rank-one conversion of reachability into mortality}
\label{conf:reduction}

We begin with the last equivalence in \cref{fig:construction-overview}. The points of the \emph{rational projective line} $\PP^1(\Q)$ correspond exactly to the one-dimensional subspaces of $\Q^2$. For arbitrary rational numbers $x,y \in \Q$ with $y\ne0$, the vectors $(x,y)^{\mathsf T}$ and $(x/y,1)^{\mathsf T}$ span the same line, which we represent by $z=x/y$. The single line not represented by any rational number in this way, which is spanned by $(1,0)^{\mathsf T}$, is taken to be represented by $\infty$. Denote by $\GL_2(\Q)$ the nonsingular rational $2\times2$ matrices. A matrix $A = \left(\begin{smallmatrix}a&b\\c&d\end{smallmatrix}\right)\in\GL_2(\Q)$ sends $(z,1)^{\mathsf T}$ to $(az+b,cz+d)^{\mathsf T}$ and therefore acts by $A(z)=(az+b)/(cz+d)$, with $A(\infty)=a/c$ and any zero denominator interpreted as $\infty$.

We call the matrices in a finite family $\mathcal G \subseteq\GL_2(\Q)$ its \emph{generators}. In a product $A_\ell\cdots A_1$, the matrix $A_1$ acts first. We write $\mathcal G^*$ for the set of all finite products, including the identity, and $\mathcal G^+$ for products of positive length. Given $\mathcal G$ together with source and target points $s,t\in\PP^1(\Q)$, the question of \emph{exact projective reachability} asks whether $W(s) = t$ for some product $W\in\mathcal G^*$.

A nonzero rank-one matrix $R=u\lambda$, with column vector $u$ spanning a line $s$ and row vector $\lambda$ having the line $t$ as the kernel, has image $s$ and kernel $t$. If a product $W$ sends $s$ to $t$, then $R$ maps $\Q^2$ onto $s$, $W$ maps $s$ onto $t$, and $R$ annihilates $t$, so $RWR = 0$. This is because $RWR = (\lambda Wu)R$, and $\lambda Wu = 0$ exactly when $W(s) = t$. Conversely, every zero product over $\mathcal G\cup\{R\}$ contains $R$, since the other generators are invertible, and the identity applied between consecutive occurrences of $R$, with $W = I$ allowed, expresses the product as $R$ times scalars $\lambda W_iu$, one of which must vanish. The lemma below restates the outer-product reduction of \citep[Lemma~6 and Proposition~4]{CassaigneEtAl2014} in projective form (see also \citep[Lemma~2]{BournezBranicky2002}, and \citep[Proposition~12]{PotapovSemukhin2017} for nonsingular integer generators with determinant $\pm1$). This rank-one conversion is also used in the NP-hardness reduction in~\citep[proof of Theorem~4]{JaaxKiefer2020}. The full proof of the lemma is given in  Appendix~\ref{app:rank-one}.

\begin{lemma}[Rank-one conversion]\label{conf:rank-one}\label{complete:rank-one}
Let $\mathcal G \subseteq \GL_2(\Q)$ be finite, and let $s,t$ be rational lines. Choose a nonzero integer column $u$ spanning $s$ and a nonzero integer row $\lambda$ with kernel $t$, and set $R = u\lambda$. Then $R$ is a nonzero integer matrix of rank one, with image $s$ and kernel $t$, and
\begin{equation}\label{conf:rank-one-equivalence}
 0 \in (\mathcal G\cup\{R\})^+ \quad \Longleftrightarrow \quad \exists W \in \mathcal G^*\; W(s) = t.
\end{equation}
\end{lemma}

To prove \cref{conf:main}, we need a fixed family $\mathcal G$ and source $s=\im R$, with only the target $t=\ker R$ varying. The following theorem constructs targets $t_\rho$ reachable from $s$ exactly when $U$ halts from $(q_1,\rho,0)$, for $\rho\in\N=\{0,1,\ldots\}$. The machine $U$ is specified in Section~\ref{conf:program-section} and the proof of the result is given in Section~\ref{sec:completion}.

\begin{theorem}[Fixed-family projective reachability]
\label{conf:reachability}\label{complete:reachability}
There exist a fixed prime $p$, a fixed finite family
$\mathcal G\subseteq\Mat_2(\Z)$ with determinants in $\{p,-p\}$,
a fixed point $s\in\Q$, and a computable sequence
$(t_\rho)_{\rho\in\N}$ in $\Q$ such that
\begin{equation}\label{conf:reachability-equivalence}
 U\text{ halts from }(q_1,\rho,0)
 \quad\Longleftrightarrow\quad
 \exists W\in\mathcal G^*\; W(s)=t_\rho.
\end{equation}
\end{theorem}

Given \cref{conf:reachability}, applying \cref{conf:rank-one} with the source $s$ and target $t_\rho$ yields a computable rank-one matrix $R_\rho$ with fixed image such that $\mathcal G\cup\{R_\rho\}$ is mortal exactly when $U$ halts, which proves \cref{conf:main} (Section~\ref{sec:completion}). The rest of Part~I proves
\cref{conf:reachability}.

\section{From two counters to a reversed one-counter program}
\label{conf:program-section}

This section describes the proof of the first equivalence in Figure 1: a fixed universal two-counter machine $U$ is simulated by a one-counter program whose single counter stores both counters as prime exponents, and the program is reversed so that its starting configuration is fixed and only its target configuration depends on the input.

\subsection{The fixed universal two-counter machine}

A \emph{configuration} of the two-counter machine $U$ is a triple $(q,c_0,c_1)$, where $q$ is the current state and, for each $i\in\{0,1\}$, $c_i\in\N$ denotes the value currently stored in counter $i$. The machine has initial state $q_1$ and halt state $q_{\mathrm h}$, which has no outgoing instruction and hence no outgoing edge. Each state apart from $q_{\mathrm h}$ has one of the following instructions, which acts on only one selected counter $i \in \{0,1\}$ and leaves the other counter unchanged.

\smallskip
\noindent\textit{(1) Increment.} The instruction replaces $c_i$ by $c_i + 1$ and moves the machine to a specified destination state $r$. For example, incrementing the first counter gives $(q,c_0,c_1) \longrightarrow (r,c_0+1,c_1)$. 

\smallskip
\noindent\textit{(2) Conditional decrement.} The instruction specifies two (possibly the same) destination states, $r_{\mathrm z}$ for the \emph{zero branch} and $r_{\mathrm p}$ for the \emph{positive branch}. If $c_i = 0$, the machine moves to $r_{\mathrm z}$ without changing either counter. If $c_i>0$, it replaces $c_i$ by $c_i-1$ and moves to $r_{\mathrm p}$. For example, a conditional decrement of the first counter gives $(q,0,c_1)\longrightarrow(r_{\mathrm z},0,c_1)$ when $c_0=0$, and $(q,c_0,c_1)\longrightarrow(r_{\mathrm p},c_0-1,c_1)$ when $c_0>0$.

We start from the two-counter simulator of Ivanov, Pelz, and Verlan \citep[Section~3]{IvanovPelzVerlan2014}, detailed in \citep[Section~5.2 and appendix]{Ivanov2015}, which simulates Korec's universal register machine \citep[Main Theorem~(b2), Definition~2.3(v)]{Korec1996}. We obtain $U$ by merging states in repeated routines of this simulator, without changing whether it halts (Appendix~\ref{counts:source-quotient}). The following proposition is proved in Appendix~\ref{prog:source-proof}.

\begin{proposition}[Fixed two-counter machine]\label{prog:source}
For the fixed machine $U$, there is no algorithm that, given $\rho\in\N$, decides whether $U$ halts from $(q_1,\rho,0)$.
\end{proposition}

\subsection{Prime-exponent simulation by a one-counter program}

We encode both counters in a single nonnegative integer $n$ using the prime-exponent simulation \citep{Minsky1967}. For a prime $\ell$, let $v_\ell(m)$ denote the exponent of $\ell$ in the positive integer $m$. Setting $m=n+1>0$, we interpret $v_2(m)$ and $v_3(m)$ as the values of counters zero and one, respectively.

\begin{definition}\label{prog:arithmetic-model}
A \emph{one-counter arithmetic program} (or simply a \emph{program}
when the meaning is unambiguous) is a finite directed graph, allowing
loops and multiple outgoing edges, whose edges are labelled by partial
maps $\N\to\N$. A configuration $(q,n)$ consists of a program state $q$
and a counter value $n\in\N$. An edge $q\xrightarrow{f}r$ permits the
step $(q,n)\to(r,f(n))$ exactly when $f(n)$ is defined. We write
$(q,n)\to^*(r,m)$ when a finite, possibly empty sequence of these steps
reaches state $r$ with counter value exactly $m$.
\end{definition}

We first construct a one-counter arithmetic program $F$ on the states of $U$, initialized at $(q_1,2^\rho-1)$ to simulate $U$ from $(q_1,\rho,0)$, since $v_2(2^\rho)=\rho$ and $v_3(2^\rho)=0$. Each transition $q\to r$ of $U$ becomes an edge $q\xrightarrow{f}r$ in $F$. We describe the updates using $m=n+1$, so an update $m\mapsto g(m)$ gives $f(n)=g(n+1)-1$. The following three cases specify the one-counter updates and their domains.

\smallskip
\noindent\textit{(1) Increments.} An increment of counter zero or one multiplies $m$ by two or three, respectively. This increases the corresponding prime exponent by one and leaves the other unchanged.

\smallskip
\noindent\textit{(2) Zero branches.} The zero branch of a conditional decrement leaves $m$ unchanged. For counter zero, it is defined when $2\nmid m$, equivalently $n\equiv0\pmod2$. For counter one, it is defined when $3\nmid m$, equivalently $n\bmod3\in\{0,1\}$.

\smallskip
\noindent\textit{(3) Positive branches.} For counter zero, the positive branch applies $m\mapsto5m/2$ when $2\mid m$. For counter one, it applies $m\mapsto5m/3$ when $3\mid m$. Division reduces the corresponding prime exponent by one, while multiplication by five changes neither exponent because five is coprime to two and three. We include this factor so that both updates increase $m$. The reversal of these instructions multiply $m$ by $2/5$ or $3/5$, which we call \emph{contractions}.

\subsection{Reachability from a fixed configuration}

The starting configuration $(q_1,2^\rho-1)$ of $F$ depends on the
input, but the fixed source line $\im R$ requires a fixed starting
configuration (\cref{conf:main,conf:rank-one}). Reversing the
transitions would put the input in the target configuration.
Since the counter may have different values at $q_{\mathrm h}$,
we first arrange for all halting computations to reach the same
final configuration.

\smallskip
\noindent\textit{(1) Create a common final configuration.} We add the loop $n\mapsto n-1$, for $n\ge1$, to $F$ as the only outgoing edge at $q_{\mathrm h}$. This enables every halting computation to reach $(q_{\mathrm h},0)$.

\smallskip
\noindent\textit{(2) Reverse the transitions.} We construct $P$ by
replacing each edge $q\xrightarrow{f}r$ of $F$ with
$r\xrightarrow{f^{-1}}q$. Each edge map is injective on its domain,
so its inverse is defined on its range.
Table~\ref{conf:instruction-table}
lists the inverse updates and their domains in $n=m-1$.

\begin{table}[!ht]
\centering
\begingroup
\normalsize
\setlength{\belowcaptionskip}{6pt}
\caption{Instructions of $P$ and the edges of $F$ they reverse, with $n,a\in\N$ and fixed $d\ge1$ and $A\subseteq\{0,\ldots,d-1\}$. Decrements refer to positive branches. Tests are identities on their domains.}
\label{conf:instruction-table}
\label{tab:instructions}
\renewcommand{\arraystretch}{1.0}
\setlength{\tabcolsep}{6pt}
\begin{tabularx}{\linewidth}{@{}l>{\raggedright\arraybackslash}X>{\raggedright\arraybackslash}Xl@{}}
\toprule
\textbf{Instruction of $P$} & \textbf{Partial update}
  & \textbf{Forward map in $F$} & \textbf{Edge of $F$}\\
\midrule
$\mathsf{Inc}$ & $n\mapsto n+1$ & $n+1\mapsto n$ & Cleanup\\
\addlinespace[2pt]
$\mathsf{Test}_{d,A}$ & $n\mapsto n$ when $n\bmod d\in A$
  & the same & Zero branch\\
\addlinespace[2pt]
$\mathsf{Div}_2$ & $2a+1\mapsto a$ & $a\mapsto2a+1$ & Increment counter zero\\
\addlinespace[2pt]
$\mathsf{Div}_3$ & $3a+2\mapsto a$ & $a\mapsto3a+2$ & Increment counter one\\
\addlinespace[2pt]
$\mathsf{Con}_2$ & $5a+4\mapsto2a+1$ & $2a+1\mapsto5a+4$ & Decrement counter zero\\
\addlinespace[2pt]
$\mathsf{Con}_3$ & $5a+4\mapsto3a+2$ & $3a+2\mapsto5a+4$ & Decrement counter one\\
\bottomrule
\end{tabularx}
\endgroup
\end{table}

The following theorem connects this construction back to $U$:
starting from $(q_{\mathrm h},0)$, $P$ reaches $(q_1,2^\rho-1)$
exactly when $U$ halts on input $\rho$. This result establishes the first equivalence in Figure~\ref{fig:construction-overview}. Its full proof is given in Appendix~\ref{prog:reversal-proof}.

\begin{theorem}\label{conf:program}\label{prog:universal}
There is a fixed one-counter arithmetic program $P$ using the instructions in \cref{conf:instruction-table} such that, for every $\rho \in \N$,
\begin{equation}\label{conf:program-equivalence}
 (q_{\mathrm h},0)\longrightarrow_P^*(q_1,2^\rho-1)
 \quad\Longleftrightarrow\quad
 U\text{ halts from }(q_1,\rho,0).
\end{equation}
\end{theorem}
\begin{proof}[Proof sketch]
Under the encoding $c_0=v_2(n+1)$ and $c_1=v_3(n+1)$, $F$ simulates the instructions of $U$ before cleanup, reproducing its counter changes, state transitions, and branch conditions. Thus $F$, started at $(q_1,2^\rho-1)$, reaches $q_{\mathrm h}$ exactly when $U$ halts on input $\rho$. Cleanup then reduces the counter until we get the  configuration $(q_{\mathrm h},0)$. Since cleanup is available only at $q_{\mathrm h}$, it cannot make a nonhalting input reach this configuration. The inverse updates defining $P$ are defined exactly on the ranges of the forward updates, so every path of $F$ reverses to a valid path of $P$, and vice versa. Therefore, the fixed program $P$ reaches $(q_1,2^\rho-1)$ from its fixed starting configuration $(q_{\mathrm h},0)$ exactly when $U$ halts on input $\rho$.
\end{proof}

\section{Encoding configurations and enforcing control with matrices}
\label{conf:control}

We represent each configuration of $P$ by a rational point representing its state and counter value. Since matrices can be multiplied in any order, we must ensure that a product maps one encoded configuration to another only when $P$ can reach the second configuration from the first one. We organize the matrix operations for each instruction in a finite directed graph, called an \emph{instruction module}. Each edge contributes one matrix to $\mathcal G$, and loops allow the number of multiplications to depend on the counter value. This section defines the encoding, the edge matrices, and the requirements for each module. Section~\ref{conf:arithmetic} constructs the modules and verifies these requirements.

\begin{definition}\label{conf:module}
An \emph{instruction module} (or \emph{module}) for a program edge
$q\xrightarrow{f}r$ is a finite directed graph with entry $q$ and exit
$r$, which may coincide. The program states $q$ and $r$ are called
\emph{boundaries}. All other vertices are \emph{intermediate states},
distinct from all program states and from the intermediate states
of every other module. Each edge starts at $q$ or an intermediate
state and ends at $r$ or an intermediate state.
\end{definition}

\subsection{Periodic counter encoding and digit checks}
\label{conf:counter-encoding}
\label{conf:digit-operations}

We encode the counter value $n$ by the rational number $z_n$, called the \emph{unary code} whose base-$p$ digits periodically repeat the block $0^n1$, where $p\ge7$ is prime. The zeros record $n$, and reading a full period returns the same code. We read digits starting with the coefficient of $p^0$. Writing $z=e+pz'$ separates the first digit $e$ from the remaining digits represented by $z'$. We call $D_e(z)=(z-e)/p$ the \emph{digit deletion map} for $e$, since applying it removes $e$ and leaves $z'$. For the block $0^n1$, the first $n+1$ digits contribute $p^n$, while the remaining digits repeat the same block. Thus $z_n=p^n+p^{n+1}z_n$, giving $z_n=p^n/(1-p^{n+1})$. A matrix word can also choose the wrong digit. To detect this, define
\begin{equation}\label{conf:valid-error-sets}
 V=\{a/b : a,b\in\Z,\ p\nmid b\} \quad\text{and}\quad E=\PP^1(\Q)\setminus V.
\end{equation} 
Every unary code $z_n$ lies in $V$ because $p$ does not divide its
denominator $1-p^{n+1}$. For a reduced fraction $z=a/b\in V$, we
compute its residue as $ab^{-1}$ modulo $p$. Applying $D_e$ gives
$(a-eb)/(pb)$, where the new factor $p$ cancels exactly when
$z\equiv e\pmod p$. Otherwise, $p$ remains in the reduced denominator
and the result lies in the \emph{error set} $E$. Every matrix in
$\mathcal G$ will keep points of $E$ in $E$, preventing any return
to an encoded configuration. The following lemma states this digit check.

\begin{lemma}\label{conf:error}
Each digit deletion $D_e$ maps $E$ into itself. For $z \in V$, we have
$D_e(z)\in V$ if and only if $z\equiv e\pmod p$.
\end{lemma}

We use this digit check in the two operations described below:

\smallskip
\noindent\textit{(1) Scan to a marker.} We call the expected digit  $e\not\equiv0\pmod p$ the \emph{marker}. A \emph{marker scan} applies $D_0$ zero or more times, then $D_e$ to delete this marker. For $z \in V$, it stays in $V$ only by deleting all initial zeros and then a digit matching $e$ modulo $p$. Thus the input fixes the scan length, including the marker deletion. For example, applying $D_0,D_0,D_1$ to $z_2$ reads $001$ and returns to $z_2$. Using $D_1$ after only one $D_0$ enters $E$, since the next digit is zero.

\smallskip
\noindent\textit{(2) Guess a length.} For division and contraction instructions  in Table~\ref{tab:instructions}, a word chooses a positive number
$b$ of zeros to delete, applies $D_0^b$, and then proceeds without
deleting the marker. On input $z_n$, every choice $1\le b\le n$
passes the digit checks and leaves $n-b$ zeros before the marker.
We call $b$ a \emph{guess} because the word chooses it before
checking whether it is correct for the instruction. For example,
division by $k$ requires the proposed output $a=n-b$ to satisfy
$n+1=k(a+1)$, but the digit checks only verify that the $b$
deleted digits are zeros.

\subsection{State encoding and matrix generators}
\label{conf:control-states}

\paragraph{Recording the state.} The unary code $z_n$
records the counter value. We encode the graph state by adding an integer chosen for that state to $z_n$. We choose these shifts so that applying a matrix intended for another state fails the digit check and sends the value into $E$. Unary codes have residue zero or one, but intermediate values in $V$ can have other residues, including after incorrect guesses. We reserve the six residue representatives $\Delta=\{-1,0,1,2,3,4\}$. Let $N$ count all program and private intermediate states in the instruction graphs. We index these states, also called \emph{phases}, by $0,\ldots,N-1$, choose a prime $p>6N$, and define the shifts and configuration encoding by
\[
 C_q(z)=z+6q+1
 \quad\text{and}\quad
 (q,n)\longmapsto C_q(z_n).
\]
During an instruction, we write the current point as $C_q(z)$,
where $z$ is the \emph{local value} obtained by removing the state
shift. The arithmetic acts on this value, which need not be a unary
code. If $z\in V$ has residue in $\Delta$, adding $6q+1$ places the
residue of $C_q(z)$ in the block $6q,\ldots,6q+5$. Since $p>6N$,
these blocks are disjoint modulo $p$. The block therefore identifies
the state $q$, and subtracting its shift $6q+1$ gives the residue
of the local value $z$.

\paragraph{Changing the counter code.} Correct digit
deletions let us read a unary code, but do not change its period
length. To change the counter value, we therefore need additional
maps, which we call \emph{auxiliary maps}. We use
$M_{\delta,h}(z)=\delta z/(1+phz)$, where $h\in\Z$ and
$\delta\in\{-1,1\}$. For $z=a/b$, this gives
$M_{\delta,h}(z)=\delta a/(b+pha)$, so these maps can change the
denominator by a multiple of the numerator. The following lemma
shows that these maps and the state shifts keep points of $E$
in $E$, so they cannot undo a failed digit check. Its proof is in
Appendix~\ref{unary:valuation-preliminaries}.
\begin{lemma}\label{conf:auxiliary-maps}
Integer translations and the maps $M_{\delta,h}$ preserve $V$ and $E$
bijectively. Moreover, $M_{\delta,h}(z)\equiv\delta z\pmod p$ for every $z\in V$.
\end{lemma}

\paragraph{Constructing each generator.} For an edge $q\to r$,
we remove the shift using $C_q^{-1}$ so that the arithmetic acts
on the local value $z$. The arithmetic constructions may need to
modify this value before or after deleting a digit, so we allow
an auxiliary map on each side of $D_e$, where $e\in\Delta$.

We choose $M_{\mathrm{pre}}=M_{1,h}$ so that the arithmetic before
deletion preserves the residue tested by $D_e$ on $V$.
After deletion, $M_{\mathrm{post}}=M_{\delta,k}$ makes any further
arithmetic change needed by the edge, and $C_r$ adds the shift
for the destination state. Here $h,k\in\Z$ and
$\delta\in\{-1,1\}$, and either auxiliary map may be the identity.
Since the rightmost factor acts first, these operations give
\begin{equation}\label{conf:generator}
 [q,r;e;M_{\mathrm{pre}},M_{\mathrm{post}}]
 =C_rM_{\mathrm{post}}D_eM_{\mathrm{pre}}C_q^{-1}.
\end{equation}
We include the product in \eqref{conf:generator} in $\mathcal G$
as one integer matrix with determinant $\delta p\in\{p,-p\}$.
Its factors are not supplied separately, so a word cannot apply
the arithmetic maps while skipping the digit check.

Suppose the current local value lies in $V$ and has residue in $\Delta$.
A generator intended for another state removes the wrong shift,
leaving a residue outside $\Delta$ that cannot match its expected
digit $e$. The map $M_{\mathrm{pre}}$ preserves this residue, so
$D_e$ sends the point into $E$, where the remaining factors keep it.
The following lemma states these state and digit checks and is
proved in Appendix~\ref{unary:valuation-preliminaries}.

\begin{lemma}\label{conf:guard}
Every generator in \eqref{conf:generator} preserves $E$.
Suppose its input is $C_{q_0}(z)$ with $z\in V$ and
$z\equiv d\pmod p$, where $d\in\Delta$.
Its output can lie in $V$ only if $q=q_0$ and $e=d$.
\end{lemma}

\subsection{Conditions for preserving program reachability}
\label{conf:module-conditions}

We now state what each module must satisfy so that matrix
words between encoded configurations follow valid computations of $P$.
For a module implementing $q\xrightarrow{f}r$ and each $n\in\N$,
consider paths from its entry state $q$ with encoded input $C_q(z_n)$, whose local value is the unary code $z_n$. A path remains in $V$
if its input and the value after every generator application lie in $V$.
It completes an instruction if it has positive length and stops at the
next program boundary it reaches. Each module must satisfy the following
three conditions.

\smallskip
\noindent\textit{(1) Residues.} Every local value along a path
remaining in $V$ must have residue in $\Delta$, including after
incorrect guesses. This keeps the state checks valid, so an
incorrect arithmetic guess cannot let a word use a generator
intended for another state to bypass the remaining tests.

\smallskip
\noindent\textit{(2) Input and output.} For each allowed input
$n\in\operatorname{dom}(f)$, we require at least one path that
completes the instruction while staying in $V$. Every such path
must give the correct local output $z_{f(n)}$. For any other input,
no path may complete the instruction without entering $E$.

\smallskip
\noindent\textit{(3) Effective construction.} We must be able to
construct the finite graph and edge-matrix formulas independently
of $n$ and before choosing $p$. The formulas may contain $p$, but
must give computable matrix entries once it is chosen. Thus one
finite matrix family works for every $n$.

\begin{theorem}\label{conf:compiler}
Suppose each instruction occurrence of a finite program has a module
satisfying the three requirements of Section~\ref{conf:module-conditions}
for every prime $p\ge7$. After listing all phases
and choosing one prime $p>6N$, their generators form a finite set
$\mathcal G\subseteq\Mat_2(\Z)$ with determinants $\pm p$ such that,
for all program configurations $(q,n)$ and $(r,k)$,
\begin{equation}\label{conf:compiler-equivalence}
 (q,n)\longrightarrow^*(r,k)
 \quad\Longleftrightarrow\quad
 \exists W\in\mathcal G^*: W(C_q(z_n))=C_r(z_k).
\end{equation}
\end{theorem}
\begin{proof}[Proof sketch]
Each allowed step $(q,n)\to_P(r,m)$ has a matrix word taking
$C_q(z_n)$ to $C_r(z_m)$. Joining these words implements the whole
computation. Conversely, suppose $W(C_q(z_n))=C_r(z_m)$ for a word
$W$ over $\mathcal G$. The point cannot enter $E$ along this word,
since every generator preserves $E$ and the target point $C_r(z_m)$ lies outside $E$.
The residue requirement keeps local residues in $\Delta$, so even
an incorrect arithmetic guess cannot make the value pass a check
for another state. Each instruction's private intermediate states
therefore prevent switching instructions to escape the remaining
tests. The input and output requirement ensures that each completed
instruction has an allowed input and the correct output. Since
$C_r(z_m)$ encodes the program state $r$, the word cannot finish halfway
through an instruction. Its completed instructions therefore give
$(q,n)\to_P^*(r,m)$. 
\end{proof}

\section{Implementing the arithmetic instructions with matrices}
\label{conf:arithmetic}

A word can pass the state and digit checks of Section~\ref{conf:control}
while guessing the wrong number of zeros to delete. The six instruction
modules in \cref{conf:instruction-table} include arithmetic tests that
reject such guesses before an instruction finishes. By
\eqref{conf:generator}, each edge $q\to r$ gives one integer matrix
in $\mathcal G$ with determinant $\pm p$. It removes the shift for $q$,
applies the digit deletion with its auxiliary maps, and adds the shift
for $r$. The auxiliary maps are not supplied separately.
The arithmetic calculations use local values
with the state shift removed.
Appendices~\ref{sec:unary} and~\ref{sec:contractions} give the full
generator formulas.

\subsection{Increment and congruence tests using digit scans}
\label{conf:basic-updates}

\noindent\textbf{Increment.} To increment the counter, we insert one zero after its marker. The
map $S(z)=z/(1-p(p-1)z)$ changes the period of $z_n$ from $0^n1$
to $0^n10$. Scanning through the marker rotates the new zero to the
front, giving
\begin{equation}\label{conf:increment-identity}
 D_1D_0^nS(z_n)=z_{n+1}.
\end{equation}

For an increment instruction from $q$ to $r$, a private state $j$
records that $S$ has already been applied. Both entry generators apply
$S$. One deletes a zero and moves to $j$, while the other deletes
the marker and moves directly to $r$ when $n=0$.
At $j$, a zero deletion stays there and a marker deletion moves
to $r$. The digit checks force all $n$ zeros to be deleted before the
marker. For example, the loop at $j$ uses the integer matrix
\[
 C_jD_0C_j^{-1}
 =\begin{pmatrix}1&(p-1)(6j+1)\\0&p\end{pmatrix}.
\]
For $z\in V$, the loop sends $C_j(z)$ to $C_j(z/p)$, keeping state
$j$ when the next digit is zero and entering $E$ otherwise.
The state shifts cancel along the path, so
\eqref{conf:increment-identity} gives a word $W$ satisfying
$W(C_q(z_n))=C_r(z_{n+1})$. All local values reached in $V$ have
residue zero or one, keeping the state checks valid
(\cref{unary:basic-instructions}).

\smallskip
\noindent\textbf{Congruence tests.} For $\mathsf{Test}_{d,A}$, we need to check whether $n\bmod d\in A$
without changing the counter. A scan of $0^n1$ returns $z_n$, while
its private states record the number of zeros modulo $d$.
At the marker, an exit is available exactly when the recorded count lies in $A$.
Otherwise, another zero deletion tries to remove the marker and enters
$E$, while the state checks prevent using another state's marker exit
(\cref{unary:congruence-tests}). 

\subsection{Division by comparing period lengths}
\label{conf:division}
\label{conf:comparison}

For $\mathsf{Div}_k$, where $k\in\{2,3\}$, the output $a$ must satisfy
$n+1=k(a+1)$, so we must divide the unary period length by $k$.
The three steps below guess an output, compare period lengths,
and recover $z_a$. We then describe the corresponding matrix products.

\paragraph{Guessing the output by deleting zeros.} A word chooses a positive
number $b$ of zeros to delete. An exit generator deletes the last chosen
zero and applies $T(z)=z/(1-pz)$. With $a=n-b$, the deletions alone
rotate the period to $0^a1\,0^b$, giving
$D_0^b(z_n)=p^a/(1-p^{n+1})$, rather than
$z_a=p^a/(1-p^{a+1})$. The first deletion rejects $n=0$, but any
$1\le b\le n$ passes the digit checks. Thus $b$ is a \emph{guess}
whose proposed output $a$ still has to satisfy $n+1=k(a+1)$.

\paragraph{Comparing period lengths by counting markers.} Applying $T$ gives
$p^a/(1-p^{a+1}-p^{n+1})$. Powers of $p^{a+1}$ alone in its expansion produce
ones separated by $a$ zeros. When $n+1=k(a+1)$, the first contribution
from $p^{n+1}$ coincides with the next marker after $k$ ones, producing
two. We therefore check for $k$ marker-one scans followed by a marker-two
scan. Each scan reads up to the next nonzero digit, so the private
states only count markers, without knowing $a$. Before the marker-two
scan, we apply $T^{-1}$, which preserves the next nonzero residue,
and negate after that scan to prepare for recovery.
\Cref{unary:common-front} proves that, for $k=2,3$, this comparison
accepts exactly when $n+1=k(a+1)$.

For example, take $n=3$ and $k=2$. The zero checks permit all three guesses
$b=1,2,3$. Choosing $b=2$ rotates $0001$ to $0100$, proposing $a=1$.
After applying $T$, the first nonzero digits are $1,1,2$, so the comparison
accepts. The other two guesses give $1,1,1,1,2$. After two marker-one
scans, the next scan expects two but encounters another one and enters $E$.

\paragraph{Recovering the unary code for the output.} Passing the comparison
verifies $a$, but we still need its unary code for the next instruction.
Applying $T$ after the comparison's final negation gives $k-1$ finite copies
of $0^a1$. Deleting $k-2$ copies leaves $p^a$, representing one block,
and $T(p^a)=z_a$ makes it repeat. Division by two needs no additional
scan, while division by three needs one.

\vspace{12pt}

For a division instruction from $q$ to $r$, the zero-deletion loop at
a private state $g$ uses the matrix $C_gD_0C_g^{-1}$. Its exit to the
first comparison state $u$ uses $C_uTD_0C_g^{-1}$, combining the final
zero deletion and $T$ in one generator as in \eqref{conf:generator}.
A one-deletion guess goes directly from $q$ to $u$, while longer guesses
pass through $g$. In either case, the intermediate shifts cancel to give
$C_uTD_0^bC_q^{-1}$.
Thus different guesses give different words over the same fixed
generators. The comparison and recovery matrices then check the guess
and produce the output (\cref{unary:odd-halving,unary:ternary-division}).
Any word $W$ completing this module while remaining in $V$ must
satisfy $n+1=k(a+1)$ and
\begin{equation}\label{conf:division-matrix-action}
 W(C_q(z_n))=C_r(z_a).
\end{equation}

\subsection{Contractions by block removal and output recovery}
\label{conf:contractions}
\label{conf:output-recovery}

The contractions $\mathsf{Con}_2$ and $\mathsf{Con}_3$ map $5a+4$ to $2a+1$
and $3a+2$, respectively. We first check $n+1=5(a+1)$ using the comparison
with $k=5$. This time, some incorrect guesses pass, and even a correct
guess leaves an extra block and marker. The four steps below reject
these errors and recover the required unary output.

\paragraph{Comparing the guessed and input period lengths.} We use the same
generator pattern as the comparison in Section~\ref{conf:division}, with
private states for five marker-one scans followed by a marker-two scan.
Deleting $b$ zeros proposes $a=n-b$, with intended choice $b=4(a+1)$.
The test counts markers without comparing the gaps between them.
For $k=5$, an incorrect guess can also give five ones followed by two,
but with different gaps.

\paragraph{Rejecting incorrect splits that pass the comparison.} Applying $T$
to a correct comparison output gives four finite copies of $0^a1$.
We remove two copies for $\mathsf{Con}_2$ and one for $\mathsf{Con}_3$.
The first marker-one scan also rejects every incorrect guess that passed
the comparison. After the leading zeros, it encounters residue $3$, $4$,
or $-1$ instead of one, so the deletion
enters $E$ (\cref{contract:front-arithmetic}). Applying $T^{-1}$ to the
remaining copies gives a periodic code with zero blocks of lengths $a$
and the required output $2a+1$ or $3a+2$ (\cref{contract:front-module}).

For example, on input $n=4$, the guess $b=3$ proposes $a=1$. It passes the
initial comparison even though $n+1\ne5(a+1)$, but the following marker-one
scan encounters residue three and rejects it. The correct guess $b=4$ gives
$a=0$ and four finite copies of the digit one. For $\mathsf{Con}_2$,
deleting two copies leaves two ones, which $T^{-1}$ recodes as the period
$1\,0\,(-1)$. This contains the required output block of one zero, but the
next instruction needs the unary period $01$.

\paragraph{Guessing the deletions needed to recover the output.} Scanning
only rotates the two blocks, so we must also change the encoding.
For $\mathsf{Con}_m$, where $m\in\{2,3\}$, a marker-one scan brings the longer
block to the front, giving
$-p^{m(a+1)-1}/(1+p^{a+1}+\cdots+p^{m(a+1)})$.
We want the denominator $1+p^{m(a+1)}$, so we must remove its $m-1$
interior powers. We make $m-1$ further guesses, each deleting a nonempty
zero prefix and applying
$T^{-1}(z)=z/(1+pz)$. The deletions lower the numerator's exponent,
and $T^{-1}$ adds $p$ times this negative numerator to the denominator,
subtracting the selected power. Choosing each deletion length to be
$a+1$ removes exactly the interior terms. A scan to the marker $-1$,
followed by two applications of $T$, then gives $z_{m(a+1)-1}$.

\paragraph{Checking the recovery guesses with marker scans.} Zero checks
do not establish that the selected powers were present. Wrong lengths
subtract at least one absent power, leaving a negative coefficient before
the final power $p^{m(a+1)}$. For $m=2,3$,
\cref{contract:projection-arithmetic} shows that the resulting candidate
has at most $m-1$ markers equal to one before a residue $2$ or $-1$.
Thus $m$ marker-one scans reject every incorrect candidate. For the
correct unary code, each scan reads one period and returns the same value.

\vspace{12pt}

For a contraction instruction from $q$ to $r$, each recovery guess uses
the same zero-deletion loop and exit construction as above, with $T^{-1}$
in place of $T$. If $v$ is the last validation state, the final exit uses
$C_rD_1C_v^{-1}$, deleting that scan's marker and moving to program state $r$.
The state checks prevent a word remaining in $V$ from skipping the
preceding scans to use this exit. Producing a candidate output
therefore does not complete the instruction. Starting from a unary input,
a word $W$ can complete the $\mathsf{Con}_m$ module while remaining in $V$
only on an input $C_q(z_{5a+4})$, and then
\begin{equation}\label{conf:contraction-matrix-action}
 W(C_q(z_{5a+4}))=C_r(z_{m(a+1)-1}).
\end{equation}

\subsection{Correct updates and residue bound}
\label{conf:instruction-verification}

An incorrect guess may pass several digit checks before it is rejected.
To apply \cref{conf:compiler}, the intervening local values must still
have residues in $\Delta$, so the word cannot switch instructions to
bypass the remaining tests.

\begin{proposition}\label{conf:instruction-implementations}
Every instruction in \cref{conf:instruction-table} has a finite
module satisfying the three requirements of Section~\ref{conf:module-conditions},
for every prime $p\ge7$. Its graph and edge-matrix formulas can be
constructed independently of the input counter and before choosing $p$.
\end{proposition}
\begin{proof}[Proof sketch]
For each allowed update $n\mapsto f(n)$, the intended guesses and scans
give a path from $z_n$ to $z_{f(n)}$. Conversely, division's comparison
forces the correct guess, and recovery returns its unary code.
For contractions, the extra scan rejects false
splits, and the final scans reject incorrect recovery lengths. Hence every
completed instruction has an allowed input and the correct output.
The appendix calculations show that local residues remain in $\Delta$
after every generator application along a path remaining in $V$, including after
incorrect guesses. This keeps the state checks valid. The same finite
graphs and edge formulas work for every prime $p\ge7$. Only loop
repetitions depend on the counter. The detailed proof is in
Appendix~\ref{contract:assembly}.
\end{proof}

\section{Completing the proof of the main result}
\label{conf:assembly}\label{sec:completion}

Let $N$ be the total number of distinct program states and private
intermediate states in the instruction graphs, and choose a prime satisfying $p>6N$. \Cref{conf:instruction-implementations,conf:compiler} give a finite family $\mathcal G$ of integer matrices with determinants $\pm p$ that ensures that whenever $P$ moves from configuration $(q,n)$ to configuration $(r,m)$ in one step, there is a product of matrices from $\mathcal G$ that maps the code $C_q(z_n)$ to the code $C_r(z_m)$. Joining these products gives a matrix product for any finite computation of $P$. Conversely, \cref{conf:compiler} shows that, if a product $W$ of
matrices from $\mathcal G$ satisfies $W(C_q(z_n))=C_r(z_m)$, then
$P$ can reach $(r,m)$ from $(q,n)$ by a finite sequence of allowed
steps. Since the instruction graphs, edge formulas, and prime $p$
are fixed independently of $\rho$, the same family works for every
input. Using the state indices from
Section~\ref{conf:control-states}, set $s=C_{q_{\mathrm h}}(z_0)$
and $t_\rho=C_{q_1}(z_{2^\rho-1})$. These points encode the fixed
starting configuration and computable target configuration of $P$
in \cref{conf:program}, so only $t_\rho$
depends on $\rho$. We now prove the main result.

\begin{proof}[Proof of \cref{conf:reachability}]
For every $\rho\in\N$,
\[
 U\text{ halts from }(q_1,\rho,0)
 \;\Longleftrightarrow\;
 (q_{\mathrm h},0)\longrightarrow_P^*(q_1,2^\rho-1)
 \;\Longleftrightarrow\;
 \exists W\in\mathcal G^*\; W(s)=t_\rho.
\]
The first equivalence follows from \cref{conf:program} and the second
from \cref{conf:compiler}, proving
\eqref{conf:reachability-equivalence}.
\end{proof}

\begin{proof}[Proof of \cref{conf:main}]
We encode the target line represented by $t_\rho$ as the kernel of a rank-one matrix with fixed image. Let $u$ be a fixed nonzero integer column spanning the line $L$ represented by $s$. Write $t_\rho=c_\rho/d_\rho$ with $c_\rho,d_\rho\in\Z$ and $d_\rho>0$, and set $R_\rho=u(d_\rho,-c_\rho)$. Then $R_\rho$ has image $L$, and its kernel is the target line because $(d_\rho,-c_\rho)(c_\rho,d_\rho)^T=0$. By \cref{conf:rank-one,conf:reachability}, $\mathcal G\cup\{R_\rho\}$ is mortal if and only if $U$ halts from $(q_1,\rho,0)$. Since $\rho\mapsto R_\rho$ is computable, this reduces the undecidable halting problem in \cref{prog:source} to mortality with $\im R_\rho=L$. The matrix $R_\rho$ need not preserve $E$, because \cref{conf:rank-one} extracts a nonsingular word from every zero product and the simulation checks apply to that word (\cref{app:rank-one-error}).
\end{proof}

\begin{proof}[Proof of \cref{conf:count}]
\Cref{tab:generator-count} lists 5,798 nonsingular generators, but different entries may represent the same matrix, so $|\mathcal G|\le5798$. To make the count exact, we reserve unused states before choosing $p$. \Cref{counts:alphabet} then adds matrices of determinant $p$ on these states until the family has exactly 5,798 distinct matrices. Since the added matrices check unused states, the first such matrix in a word starting from an encoded configuration sends the value into $E$. All nonsingular generators preserve $E$, so the additions leave reachability unchanged. With this choice of $p$ and $\mathcal G$, the preceding rank-one construction adds the singular matrix $R_\rho$, giving exactly 5,799 distinct matrices.
\end{proof}

\label{main-text-end}
\clearpage
\appendix
\addtocontents{toc}{\protect\setcounter{tocdepth}{1}}
\addcontentsline{toc}{part}{Part II. Detailed proofs and constructions}
\begin{center}\large\bfseries Part II. Detailed proofs and constructions\end{center}
\vspace{-0.5\baselineskip}
\section{Proof of the rank-one conversion lemma}
\label{app:rank-one}

\begin{proof}[Proof of \cref{conf:rank-one}]
Clearing denominators gives the required integer $u$ and $\lambda$,
and $R=u\lambda$ has the stated rank, image, and kernel.
For every $W\in\mathcal G^*$, we have $RWR=(\lambda Wu)R$,
so $W(s)=t$ gives a zero product, including $R^2$ when $W=I$.
Conversely, every zero product contains $R$, since the other
generators are invertible. Remove the invertible factors before
the first and after the last occurrence of $R$. Repeated use of
the same identity expresses the remaining product as $R$
multiplied by the scalars $\lambda Wu$ for the products
$W\in\mathcal G^*$ between consecutive occurrences of $R$.
Since $R\ne0$, at least one of these scalars is zero, so the
corresponding $W$ satisfies $W(s)=t$.
\end{proof}

\begin{remark}\label{app:rank-one-error}
The rank-one matrix $R_\rho$ in the proof of \cref{conf:main} need
not preserve the error set $E$ of Section~\ref{conf:counter-encoding}.
That requirement applies only to the nonsingular simulation matrices.
Even when a zero product contains many copies of $R_\rho$,
\cref{conf:rank-one} extracts a product of nonsingular generators to
which \cref{conf:compiler} applies.
\end{remark}

\section{The fixed machine and reversed program}
\label{sec:programs}

We obtain $U$ by merging states in repeated routines of the two-counter simulator
$U_2$ of Ivanov, Pelz, and Verlan \citep[Section~3]{IvanovPelzVerlan2014}.
This preserves whether the machine halts from the same initial counter values
and reduces the generator count. We use the increment and conditional-decrement
instructions defined in Section~\ref{conf:program-section}, following
Ivanov's deterministic register-machine model
\citep[Section~2.2, pp.~28--29]{Ivanov2015}.

\subsection{State identifications in the two-counter simulator}
\label{counts:source-quotient}

The machine $U_2$ has states $q_1,\ldots,q_{278}$, initial
state $q_1$, and halt state $q_{278}$. Its instruction list
\citep[appendix, pp.~128--129]{Ivanov2015} contains 92 increments
of counter zero, 73 increments of counter one, 80 conditional
decrements of counter zero, and 32 conditional decrements of
counter one. We identify the states listed in \cref{tab:source-quotient}.

\begin{table}[htbp]
\centering
\begingroup
\normalsize
\setlength{\belowcaptionskip}{6pt}
\caption{Each row lists three states of $U_2$ that are merged into one.
State numbers abbreviate their $q$ labels, and all unlisted states remain
distinct. Here $\operatorname{INC}_c$ and $\operatorname{DEC}_c$ denote
an increment and a conditional decrement of counter $c$.
Entries in the third column correspond in order to the states in the first.}
\label{tab:source-quotient}
\renewcommand{\arraystretch}{1.1}
\setlength{\tabcolsep}{6pt}
\begin{tabularx}{\linewidth}{@{}ll>{\raggedright\arraybackslash}Xl@{}}
\toprule
\textbf{States to identify} & \textbf{Operation}
 & \textbf{Positive/increment destinations} & \textbf{Zero destination}\\
\midrule
$45,150,270$ & $\operatorname{DEC}_1$ & $46,151,271$ & $1$\\
\addlinespace[3pt]
$46,151,271$ & $\operatorname{INC}_0$ & $45,150,270$ & ---\\
\addlinespace[3pt]
$200,230,256$ & $\operatorname{DEC}_1$ & $201,231,257$ & $258$\\
\addlinespace[3pt]
$201,231,257$ & $\operatorname{INC}_0$ & $200,230,256$ & ---\\
\bottomrule
\end{tabularx}
\endgroup
\end{table}

Let $\pi$ map each state of $U_2$ to its merged state, with one merged
state for each row of \cref{tab:source-quotient} and every unlisted
state left unchanged. To construct $U$, we replace each transition
$q\to r$ of $U_2$ by $\pi(q)\to\pi(r)$, keeping its counter operation
and branch condition. The initial state $q_1$ and halt state
$q_{\mathrm h}=q_{278}$ are not merged with any other state.

\begin{lemma}\label{prog:quotient}
The machine $U$ is well defined. For every $u,v\in\N$, it halts
from $(q_1,u,v)$ exactly when $U_2$ does. Its instruction counts are
\begin{equation}\label{prog:source-counts}
 (I_0,I_1,J_0,J_1)=(88,73,80,28),
\end{equation}
where $I_c$ and $J_c$ count increments and conditional decrements
of counter $c$.
\end{lemma}

\begin{proof}
The states in each merged group have the same instruction type and
selected counter. Their corresponding destinations also map to the
same state under $\pi$, so each merged state has one well-defined
instruction. Start both machines at $q_1$ with the same counter values.
Whenever $U_2$ is at $(q,c_0,c_1)$ and $U$ is at
$(\pi(q),c_0,c_1)$, with $q\ne q_{278}$, they perform the same counter operation, choose
the same branch when a test is required, and reach states related by
$\pi$. Induction preserves this relation after every step.
Since only $q_{278}$ maps to the halt state, the two machines either
both run forever or halt after the same number of steps.
Each row of the table replaces three instructions
by one. The two increment rows therefore remove four increments of
counter zero, and the two decrement rows remove four conditional
decrements of counter one, giving \cref{prog:source-counts}.
\end{proof}

\subsection{Undecidability of halting from an encoded input}
\label{prog:source-proof}

\begin{proof}[Proof of \cref{prog:source}]
Korec's fixed register machine $U_{20}$ is universal
\citep[Main Theorem~(b2), p.~268]{Korec1996}. A register machine
and its input can therefore be encoded computably into the initial
register values of $U_{20}$. In Korec's definition, after encoding
the input, $U_{20}$ followed by a total computable output decoder
computes the same partial function as the encoded machine
\citep[Definition~2.3(v) and the
following paragraph, pp.~275--276]{Korec1996}. Since the decoder is
defined on every output, the decoded result exists exactly when
$U_{20}$ reaches its designated halt state. Equality of the partial
functions therefore means that $U_{20}$ reaches that state exactly
when the encoded machine halts.

The two-counter machine $U_2$ simulates $U_{20}$
\citep[Section~3]{IvanovPelzVerlan2014}, using the prime-exponent
encoding of \citep{Minsky1967}. The construction is detailed in
\citep[Section~5.2, pp.~98--99]{Ivanov2015}. If
$(a_1,\ldots,a_k)$ are the initial values of the $k$ registers of
$U_{20}$, then $U_2$ starts with
$\rho=\prod_{i=1}^k\ell_i^{a_i}$ in counter zero and zero in counter
one, where the $\ell_i$ are fixed distinct primes. Counter one holds
intermediate results. Each simulated instruction is implemented by
a terminating arithmetic routine that returns counter one to zero.
Thus $U_2$ reaches its unique halt state exactly when $U_{20}$ reaches
its designated halt state. Composing the input encodings gives a
computable reduction from the halting problem for register machines
to whether $U_2$ halts from $(q_1,\rho,0)$.
By \cref{prog:quotient}, the same equivalence holds for $U$.
\end{proof}

\subsection{Proof of the program equivalence theorem}
\label{prog:reversal-proof}

\begin{proof}[Proof of \cref{conf:program}]
Fix $\rho\in\N$. We interpret a configuration $(q,n)$ of $F$ as
the configuration $(q,v_2(n+1),v_3(n+1))$ of $U$. In particular,
$(q_1,2^\rho-1)$ represents $(q_1,\rho,0)$.

\paragraph{Simulating the instructions of $U$.}
Let $q\ne q_{\mathrm h}$ and put $m=n+1>0$,
$c_0=v_2(m)$, and $c_1=v_3(m)$. In $F$, the edges simulating
increments of counters zero and one multiply
$m$ by two or three, increasing $c_0$ or $c_1$, respectively,
by one and leaving the other exponent unchanged. Both are defined
for every $n\in\N$.

In $F$, the zero branch of a conditional decrement leaves $m$ unchanged and
is defined when $2\nmid m$ for counter zero or $3\nmid m$ for
counter one, exactly when the selected counter has value zero.
Otherwise, the positive branch multiplies $m$ by $5/2$ or $5/3$,
respectively. Since five is coprime to two and three, these give
$v_2(5m/2)=c_0-1$ and $v_3(5m/2)=c_1$, or
$v_2(5m/3)=c_0$ and $v_3(5m/3)=c_1-1$.
Both positive branches keep $m$ a positive integer, so $n=m-1$
remains nonnegative. For each conditional decrement, exactly one
of its two branches is defined. Each edge has the same destination
state as the corresponding branch in $U$.

Together with the increment cases, these checks show that exactly
one outgoing edge of $F$ is defined whenever $q\ne q_{\mathrm h}$,
and it represents the step prescribed by $U$.
Induction on the number of steps therefore shows that every finite
path of $F$ from $(q_1,2^\rho-1)$ containing no cleanup step
represents a finite computation of $U$ from $(q_1,\rho,0)$,
and each such computation of $U$ has a corresponding path of $F$.

\paragraph{Reaching a common final configuration.}
The only outgoing edge of $F$ at $q_{\mathrm h}$ is cleanup,
$n\mapsto n-1$ for $n\ge1$. If $U$ halts, $F$
reaches some $(q_{\mathrm h},n)$, and exactly $n$ cleanup steps
reach $(q_{\mathrm h},0)$, including zero steps when $n=0$.
Conversely, any path of $F$ from $(q_1,2^\rho-1)$ to
$(q_{\mathrm h},0)$ has a first visit to $q_{\mathrm h}$.
Its prefix up to that visit contains no cleanup step and therefore
represents a halting computation of $U$. We have proved
\[
 (q_1,2^\rho-1)\longrightarrow_F^*(q_{\mathrm h},0)
 \quad\Longleftrightarrow\quad
 U\text{ halts from }(q_1,\rho,0).
\]

\paragraph{Checking the inverse maps and their domains.}
Each inverse must be defined exactly on the range of its forward map,
so that $P$ can reverse only steps that $F$ could take.
We check that the edge maps of $F$ are injective and that their inverses
are the corresponding instructions in Table~\ref{conf:instruction-table}.
In the following formulas,
$a\in\N$. Cleanup maps $a+1$ to $a$, so its range is all of
$\N$ and its inverse is $\mathsf{Inc}$. A zero branch is the
identity on its domain, so reversing it gives the same test,
$\mathsf{Test}_{2,\{0\}}$ or $\mathsf{Test}_{3,\{0,1\}}$.
In $F$, the edges simulating increments of counters zero and one
send the one-counter value $a$ to $2a+1$ and $3a+2$, respectively.
Each map is injective, with range exactly the domain of
$\mathsf{Div}_2$ or $\mathsf{Div}_3$, and these division instructions
recover $a$.

For the edge of $F$ simulating a positive decrement of counter zero,
the allowed inputs are exactly $n=2a+1$. Then $m=2(a+1)$, so
multiplying $m$ by $5/2$ gives the new one-counter value
$5(a+1)-1=5a+4$. The edge simulating a positive decrement of counter
one has allowed inputs $n=3a+2$. Multiplying $m=3(a+1)$ by $5/3$
gives the same output $5a+4$.
Both maps are injective, and both have range exactly the nonnegative
integers congruent to four modulo five. Their inverses are therefore
$\mathsf{Con}_2(5a+4)=2a+1$ and
$\mathsf{Con}_3(5a+4)=3a+2$, with the domains stated in the table.

\paragraph{Reversing paths and fixing the starting configuration.}
The domain checks above show that each edge $r\xrightarrow{f^{-1}}q$
reverses precisely the allowed steps of $q\xrightarrow{f}r$.
Reversing the order of a finite path of $F$ therefore gives a valid
path of $P$, and conversely. Several reversed edges may be available
at one configuration, but each still reverses an allowed step of $F$.
Together with the cleanup equivalence above, this proves
\eqref{conf:program-equivalence}.
The states, edge maps, and domains of $P$ depend only on the fixed
machine $U$. Thus $P$ and its starting configuration
$(q_{\mathrm h},0)$ are fixed. Only the target configuration
$(q_1,2^\rho-1)$ depends on $\rho$, and the map
$\rho\mapsto2^\rho-1$ is computable.
\end{proof}

\section{Valuations, scans, and elementary instruction modules}\label{sec:unary}

This appendix proves \cref{conf:error,conf:auxiliary-maps,conf:guard,conf:compiler}
and gives the arithmetic calculations and generator tables for increments,
congruence tests, and division. The arithmetic formulas use local values,
with the state shifts removed. Each table entry denotes one matrix from
\eqref{conf:generator}, with $\mathrm{id}$ denoting the identity.
Throughout, $p\ge7$ is prime. When applying \cref{conf:guard}, we also
require $p>6N$, where $N$ counts all program and intermediate states.
The intermediate states are private to their modules, as in
\cref{conf:module}. When two stages of one instruction meet at an
intermediate state, the first must leave the local value in the form
required by the second.

Several auxiliary operations on one side of a digit deletion can be
combined into one allowed map, since
$M_{\delta_2,h_2}M_{\delta_1,h_1}
=M_{\delta_2\delta_1,h_1+\delta_1h_2}$.
In particular, negation is $\mathcal N(z)=-z=M_{-1,0}(z)$.

\subsection{Encoding and program reachability}
\label{unary:valuation-preliminaries}

For a block of integer digits $e_0,\ldots,e_{L-1}$ with $L\ge1$, write
$z=\sum_{i=0}^{L-1}e_i p^i+p^Lz'$, where $z'$ represents the
remaining digits. We read the digits in increasing order of their
powers of $p$. For a periodically repeated block
$w=e_0\cdots e_{L-1}$, write $z_w$ for its rational code.
The remainder is again $z_w$, giving
$z_w=(\sum_{i=0}^{L-1}e_i p^i)/(1-p^L)$.
Deleting its first digit rotates the block, since
$D_{e_0}(z_w)=z_{e_1\cdots e_{L-1}e_0}$, and a full scan gives
$D_{e_{L-1}}\cdots D_{e_0}(z_w)=z_w$.
These algebraic identities also hold for negative digits.

We use the valuation to count initial zeros and detect failed checks.
For nonzero rational $z$, let $v_p(z)$ be the exponent of $p$ in its
numerator minus that in its denominator, and set $v_p(0)=+\infty$.
Thus $V=\{z\in\Q:v_p(z)\ge0\}$. The denominator of $z_n$ is
coprime to $p$, so $v_p(z_n)=n$.
For rational $x,y$ and integers $k\ge1$, write
$x\equiv y\pmod{p^k}$ when $v_p(x-y)\ge k$.
For rational $x,y$, we use $v_p(xy)=v_p(x)+v_p(y)$ and
$v_p(x+y)\ge\min\{v_p(x),v_p(y)\}$.
For nonzero $x,y$ with different valuations, the second relation is
an equality. Indeed, after factoring out the smaller power of $p$,
exactly one remaining term is divisible by $p$.
The point $\infty$ belongs to $E$ but has no assigned valuation.

\begin{proof}[Proof of \cref{conf:error}]
If $v_p(z)<0$, subtracting an integer leaves its valuation unchanged,
so $v_p(D_e(z))=v_p(z)-1<0$. Also $D_e(\infty)=\infty$.
For $z\in V$, the quotient $D_e(z)=(z-e)/p$ lies in $V$ exactly
when $v_p(z-e)\ge1$, equivalently when $z\equiv e\pmod p$.
\end{proof}

Write $0^*e$ for a marker scan with $e\not\equiv0\pmod p$, counting
the final marker deletion in its length. For nonzero $z\in V$, the
first $v_p(z)$ zero deletions remain in $V$ and leave
$p^{-v_p(z)}z$. We call its residue the \emph{leading residue} of
$z$. The final deletion $D_e$ remains in $V$ exactly when $e$ has
this residue. A completed scan that remains in $V$ therefore has
length $v_p(z)+1$.
Applying $D_e$ earlier or $D_0$ once more enters $E$.
At $z=0$, $D_0$ leaves the value unchanged and every such $D_e$
enters $E$.

\begin{proof}[Proof of \cref{conf:auxiliary-maps}]
For $z\in V$, the denominator $1+phz$ has residue one, so
$M_{\delta,h}(z)\in V$ and $M_{\delta,h}(z)\equiv\delta z\pmod p$.
Its inverse is $z\mapsto\delta z/(1-\delta phz)$, which also maps
$V$ into $V$. Integer translations and their inverses do the same.
Thus each map sends $V$ onto $V$. Since each is a bijection of
$\PP^1(\Q)$, it also sends the complement $E=\PP^1(\Q)\setminus V$
onto itself. This includes the poles of the maps and the point $\infty$.
\end{proof}

\begin{proof}[Proof of \cref{conf:guard}]
Each factor in \eqref{conf:generator} maps $E$ into $E$ by
\cref{conf:error,conf:auxiliary-maps}. For the input $C_{q_0}(z)$
in the lemma, removing the shift for $q$ gives
$C_q^{-1}C_{q_0}(z)=z+6(q_0-q)$. Since $M_{\mathrm{pre}}$ preserves
residues, deletion stays in $V$ exactly when
$d+6(q_0-q)\equiv e\pmod p$, or equivalently
$6q_0+1+d\equiv6q+1+e\pmod p$.
These two integers lie in the residue blocks for $q_0$ and $q$.
The blocks are disjoint because $p>6N$, so $q=q_0$, and their
distinct representatives then give $d=e$.
Otherwise deletion sends the point into $E$, where the remaining
maps keep it.
\end{proof}

\phantomsection
\label{unary:compiler-proof}
\begin{proof}[Proof of \cref{conf:compiler}]
We use the residue, input and output, and effective construction
requirements stated in Section~\ref{conf:module-conditions}.

\paragraph{Constructing one matrix family.}
The effective construction requirement lets us construct the finite
instruction graphs and their edge formulas before choosing $p$.
Use distinct intermediate states for different instruction modules,
but keep the same program state labels wherever their boundaries meet.
Let $N$ be the total number of states and index them by $0,\ldots,N-1$.
Choose a prime $p>6N$ with $p\ge7$ and evaluate each edge formula
\eqref{conf:generator}. All its factors have integer entries. Its shift matrices
have determinant one, its digit-deletion matrix has determinant $p$,
and its auxiliary matrices have determinants in $\{1,-1\}$.
Thus every edge gives an integer matrix with determinant $\pm p$.
Let $\mathcal G$ be the finite set of these matrices. Several edges
may give the same matrix, so we will not assume that a factor in a
matrix word identifies a unique edge.

The blocks $\{6s,\ldots,6s+5\}$ for state indices $s$ are disjoint
modulo $p$. If $u\in V$ has residue in $\Delta$, then $C_s(u)$
has residue in the block for $s$. In particular, the configuration
encoding is injective. Indeed, $C_q(z_n)=C_r(z_k)$ first gives
$q=r$ from the disjoint blocks, since each unary code has residue
zero or one. Cancelling the equal shifts gives $z_n=z_k$, and
$v_p(z_n)=n$ and $v_p(z_k)=k$ then give $n=k$.

\paragraph{Implementing a program computation.}
For an allowed step $(q,n)\longrightarrow(r,f(n))$, the input and
output requirement in Section~\ref{conf:module-conditions} gives a
positive-length path through its module that remains in $V$ and has
local output $z_{f(n)}$. The intermediate shifts cancel between
consecutive edge matrices, so the product for this path maps
$C_q(z_n)$ to $C_r(z_{f(n)})$. Concatenating these products along
a finite program path proves the forward implication in
\eqref{conf:compiler-equivalence}. A program path of length zero
is implemented by the identity matrix, which belongs to $\mathcal G^*$.

\paragraph{Following an arbitrary matrix word.}
For the converse, first let $W=A_\ell\cdots A_1$ have positive
length and satisfy $W(C_q(z_n))=C_r(z_k)$. Set $x_0=C_q(z_n)$
and $x_i=A_i(x_{i-1})$ for $1\le i\le\ell$, so that
$x_\ell=C_r(z_k)$. These are the points acted on by the matrices,
rather than the local values obtained by removing a state shift.
Every $x_i$ lies in $V$. Otherwise, preservation
of $E$ by \cref{conf:guard} would force all subsequent values,
including $x_\ell$, into $E$. For each factor $A_i$, choose an
edge whose matrix equals $A_i$. Such an edge exists by the
definition of $\mathcal G$.

We identify program steps successively along this word. Suppose the
first $j$ factors implement a program path ending at $(a,m)$, so
$x_j=C_a(z_m)$, and $j<\ell$. Initially this holds
with $j=0$, $a=q$, and $m=n$. Since $z_m$ has residue zero or one,
\cref{conf:guard} forces the chosen edge for $A_{j+1}$ to leave
$a$ and check the matching digit. An edge leaving a program boundary
starts a module with entry state $a$.

Suppose the edges selected so far have followed this module to an
intermediate state $s$, with current point $x_i=C_s(u_i)$ and local
value $u_i=C_s^{-1}(x_i)$. Integer translations preserve $V$, so
$u_i\in V$, and all earlier
local values on this module path also lie in $V$. The residue
requirement in Section~\ref{conf:module-conditions} therefore gives
$u_i\equiv d\pmod p$ for some $d\in\Delta$, including when the path contains an
incorrect arithmetic guess. If $i<\ell$, the next value $x_{i+1}$
is in $V$, so \cref{conf:guard} forces the chosen edge for
$A_{i+1}$ to leave $s$ with the matching digit. Since $s$ is
private to the current module, this edge belongs to the same module.
This proves inductively that the chosen edges stay in that module
until the next program boundary or the end of the word. The residue
requirement is used only after the preceding edges have been shown
to form a path in the module.

When a positive-length segment first reaches a program boundary $b$,
it completes the current instruction $a\xrightarrow{f}b$ in the
sense of Section~\ref{conf:module-conditions}. The input and output
requirement gives $m\in\operatorname{dom}(f)$ and local output
$z_{f(m)}$. Hence this segment implements the allowed program step
$(a,m)\longrightarrow(b,f(m))$, and we repeat the argument from
the resulting encoded configuration. This also covers instructions
with the same entry and exit, since completion is defined by the
first return to a program boundary after a positive number of edges.
At least one matrix factor is used per completed instruction,
so the procedure identifies finitely many program steps followed,
possibly, by a final unfinished instruction. None of this reasoning
requires the chosen edge for a matrix factor to be unique.

\paragraph{Identifying the final configuration.}
The word cannot end inside an unfinished instruction. If it did,
$x_\ell=C_s(u)$ for a private intermediate state $s$ and a local
value $u\in V$ with residue in $\Delta$, by the residue requirement.
Its residue would lie in the block for $s$. But
$x_\ell=C_r(z_k)$ lies in the block for the program state $r$,
which is disjoint from the block for $s$. Thus the entire word
splits into completed, allowed instruction steps. Injectivity of
the configuration encoding identifies the last state and counter
as $(r,k)$, proving the reverse implication in
\eqref{conf:compiler-equivalence} for nonempty words.

If the word is empty, $C_q(z_n)=C_r(z_k)$. The same injectivity
argument gives $(q,n)=(r,k)$, and the program path of length zero
establishes the reverse implication in this case as well.
\end{proof}

\subsection{The increment module}

\begin{figure}[!htbp]
  \centering
  \includegraphics[width=0.9\linewidth]{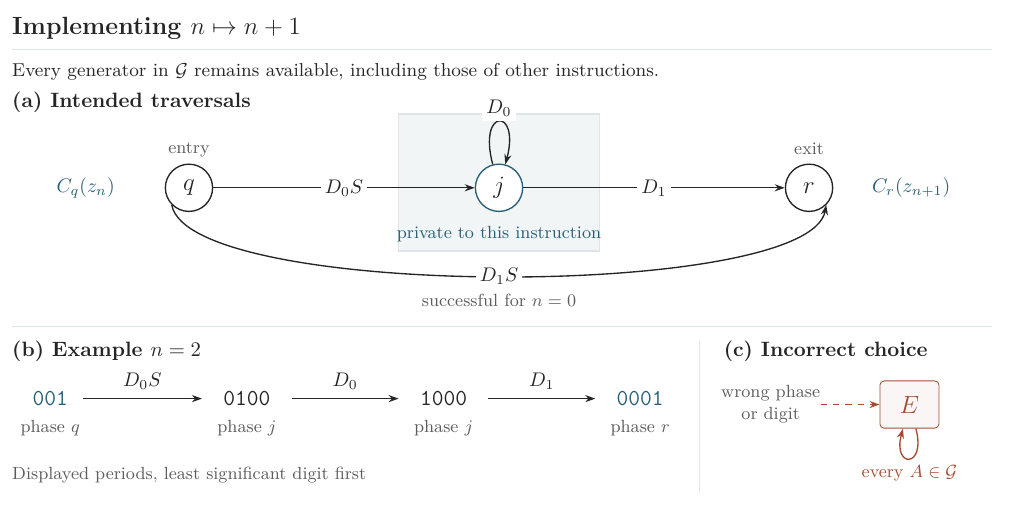}
  \caption{The increment module of \cref{unary:basic-instructions}.
    (a) Entry $q$, private intermediate state $j$, and exit $r$, with each edge
    labelled by the maps composed in its generator. (b) The local
    periods for $n=2$, least significant digit first. (c) A generator
    with the wrong state or digit sends the point into $E$, which
    every generator preserves.}
  \label{fig:increment-module}
\end{figure}

\begin{proposition}
\label{unary:basic-instructions}
A module implements exactly $\mathsf{Inc}(n)=n+1$ on the unary
codes. It has four listed generators.
\end{proposition}

\begin{proof}
The map $S=M_{1,1-p}$ adds one zero after the unary marker, giving
$S(z_n)=p^n/(1-p^{n+2})$ with period $0^n10$. Scanning through its marker gives
\eqref{conf:increment-identity}, including $n=0$.
For an $\mathsf{Inc}$ instruction from boundary $q$ to boundary $r$, reserve a private
intermediate state $j$ and supply exactly these four generators
(\cref{fig:increment-module}).
\begin{equation}\label{unary:increment-table}
 \begin{array}{ll}
 {}[q,j;0;S,\mathrm{id}],&[q,r;1;S,\mathrm{id}],\\
 {}[j,j;0;\mathrm{id},\mathrm{id}],&[j,r;1;\mathrm{id},\mathrm{id}].
 \end{array}
\end{equation}
For $n>0$, the $q\to j$ generator leaves the local value
$p^{n-1}/(1-p^{n+2})$ at $j$. To complete the instruction while
remaining in $V$, a word must make $n-1$ more zero deletions before
taking the marker exit $j\to r$.
For $n=0$, the direct $q\to r$ generator completes the increment immediately.
These scans complete the instruction while remaining in $V$ for every
$n\ge0$, with output $z_{n+1}$. The local residues stay in
$\{0,1\}$, and $S$ preserves residues, so the state and digit
checks reject any generator with the wrong state or expected digit.
\end{proof}

\subsection{Congruence-test modules}

The next construction tests the counter value by scanning a complete
unary period without changing the code. Its intermediate states record
the number of deleted zeros modulo $d$. The state for remainder zero
is still an intermediate state, so counting a multiple of $d$ zeros
cannot finish the instruction before the marker is deleted.

\begin{lemma}\label{unary:congruence-tests}
Fix an integer $d\ge1$ and a set
$A\subseteq\{0,\ldots,d-1\}$. A finite module implements
$\mathsf{Test}_{d,A}$, the identity on precisely the unary inputs satisfying
$n\bmod d\in A$. It has $1+d+|A|+\mathbf 1_{\{0\in A\}}$ listed generators. Here
$\mathbf 1_{\{0\in A\}}$
is one when $0\in A$ and zero otherwise.
\end{lemma}

\begin{proof}
For an instruction from boundary $q$ to boundary $r$, reserve private
intermediate states $a_0,\ldots,a_{d-1}$, where $a_j$ records that
the number of deleted zeros is congruent to $j$ modulo $d$.
Indices are reduced modulo $d$. Supply exactly the following generators.
\begin{equation}\label{unary:congruence-table}
 \begin{array}{lll}
 {}[q,a_{1\bmod d};0;\mathrm{id},\mathrm{id}],
    &&\text{always},\\
 {}[a_j,a_{(j+1)\bmod d};0;\mathrm{id},\mathrm{id}],
    &&0\le j<d,\\
 {}[a_j,r;1;\mathrm{id},\mathrm{id}],
    &&j\in A,\\
 {}[q,r;1;\mathrm{id},\mathrm{id}],
    &&\text{if }0\in A.
 \end{array}
\end{equation}
For $n>0$, after $b$ zero deletions, $1\le b\le n$, the
state is $a_{b\bmod d}$ and the local value is
$p^{n-b}/(1-p^{n+1})$. At $b=n$, this value has residue $1$,
so the next deletion must remove the marker.
The marker exit exists exactly when $n\bmod d\in A$ and returns $z_n$
by the full-period identity in Appendix~\ref{unary:valuation-preliminaries}.
For $n=0$, the direct $q\to r$ marker generator completes the test
exactly when $0\in A$. Otherwise, no first step remains in $V$.
If a required marker exit is absent, an attempted zero deletion enters $E$.

Every local value along a path remaining in $V$ has residue $0$ or $1$,
so the state and digit checks reject generators for other states or digits.
A full scan exists for every permitted
input, and every path completing the instruction while remaining in $V$
has unchanged output.
\end{proof}

\subsection{Length comparison and intermediate codes}
\label{unary:comparison-proof}

We prove the length comparison from Section~\ref{conf:arithmetic},
with $k=2,3$ for division and $k=5$ for contractions, including its
intermediate output and the residue bounds for incorrect guesses. Starting from $z_n$,
choose $1\le b\le n$ zero deletions and set $a=n-b$ and $Q=a+1$.
The required period relation $n+1=kQ$ is equivalent to $b=(k-1)Q$.

To retain the lengths needed for recovery, define the intermediate code
\begin{equation}\label{conf:two-field}
 X_{c,d}=\frac{p^c(1-p^{d+1})}{1-p^{c+d+2}},
\end{equation}
for $c,d\in\N$. It represents the period $0^c1\,0^d(-1)$, whose zero
blocks have lengths $c,d$. The code determines both lengths, since
its valuation gives $c$ and the formula is injective in $p^{d+1}$
for fixed $c$. The intended split will be stored as $X_{a,b-1}$.

The comparison uses $T(z)=z/(1-pz)=M_{1,-1}(z)$ and its inverse
$T^{-1}(z)=z/(1+pz)=M_{1,1}(z)$. Their denominators have residue
one, so both preserve valuations and leading residues on nonzero
elements of $V$. For the calculation, we rescale the local value as
$\tau=pz$. A marker scan of length $s$, including its
final deletion, acts by
$D_eD_0^{s-1}(z)=(z-ep^{s-1})/p^s$ and becomes
\begin{equation}\label{conf:scan-rule}
 \tau\longmapsto\frac{\tau}{p^s}-e.
\end{equation}
The marker contributes the constant $-e$ independently of the scan
length, and $T^{\pm1}$ acts by $\tau\mapsto\tau/(1\mp\tau)$.
Multiplication by $p$ shifts digit positions by one without changing
the order of nonzero digits. Thus, for nonzero $z\in V$, a completed
marker scan that stays in $V$ has length $v_p(z)+1=v_p(\tau)$.

After the guessed $b$ zero deletions, the local value is
$p^a/(1-p^{n+1})$. Applying $T$ gives
\begin{equation}\label{conf:comparison-start}
 \tau_0=\frac{p^Q}{1-p^Q-p^{Q+b}}.
\end{equation}
The following procedure tests whether exactly $k$ markers equal to
one precede the first marker equal to two. Its application of
$T^{-1}$ before the marker-two scan preserves the next leading residue.

\begin{lemma}\label{unary:common-front}
Let $p\ge7$ be prime, $k\in\{2,3,5\}$, and $n\ge1$. Choose
$1\le b\le n$, put $a=n-b$ and $Q=a+1$, and start from $z_n$.
Delete $b$ zeros and apply $T$. Then perform $k$ marker-one scans,
apply $T^{-1}$, perform a marker-two scan, and negate.
This procedure completes without entering $E$ exactly when
$b=(k-1)Q$ or ($k=5$ and $Q<b<2Q$).
For $b=(k-1)Q$, its output is $x=X_{a,(k-1)Q-1}$.
Every local value reached before a rejecting deletion or the final
marker-two deletion has residue in $\{0,1,2\}$, including those
of rejected choices of $b$.
\end{lemma}

\begin{proof}
We use \eqref{conf:comparison-start} to locate the first digit two
and count the preceding ones. The scan rule \eqref{conf:scan-rule}
then determines which guesses pass the comparison.
We find the initial digits of $\tau_0$ by congruences modulo powers of $p$. For every
integer $M\ge0$, the geometric identity gives
\begin{equation}\label{conf:comparison-expansion}
 \tau_0\equiv p^Q\sum_{j=0}^{M}(p^Q+p^{Q+b})^j
       \pmod {p^{(M+2)Q}}.
\end{equation}
Indeed, the omitted rational remainder has valuation at least
$(M+2)Q$. Each term on the right consists of an initial $p^Q$
followed by factors selected from $p^Q$ and $p^{Q+b}$.
The second exponent $Q+b=n+1$ is the original unary period length.
Products containing only $p^Q$ give coefficient $1$ at exponents
$Q,2Q,3Q,\ldots$. The term $p^Qp^{Q+b}$ gives another
coefficient $1$ at exponent $2Q+b$.
All other products involving that factor have exponent
at least $3Q+b$, where the two possible orders of one factor
$p^{Q+b}$ and one factor $p^Q$ both contribute.

The first coefficient $2$ occurs at the least exponent receiving
more than one contribution. If $b=hQ$ for an integer $h\ge1$,
this exponent is $2Q+b=(h+2)Q$. It is preceded by
$h+1$ nonzero coefficients, all equal to $1$.

If $hQ<b<(h+1)Q$ for an integer $h\ge0$, the first coefficient
$2$ occurs at exponent $3Q+b$. Before it, the $h+3$
multiples of $Q$ and the exponent $2Q+b$ give $h+4$ coefficients
equal to $1$. Through the first coefficient $2$, these coefficients
are at most $2$, so these congruences identify them with the
initial digits without carries. This classification uses no later
coefficients.

Each marker-one scan removes the next digit $1$. The subsequent
$T^{-1}$ preserves the valuation and leading residue, so the
marker-two scan finishes without entering $E$ exactly when those $k$ digits are followed
by the first $2$. In the divisible case this requires $k=h+1$,
giving $b=(k-1)Q$. In the other case it requires $k=h+4$.
Among $k\in\{2,3,5\}$, the latter case is possible only for $k=5$
and $h=1$. Thus the choices completing the comparison are exactly
\begin{equation}\label{conf:comparison-choices}
 b=(k-1)Q
 \quad\text{or}\quad
 \bigl(k=5\text{ and }Q<b<2Q\bigr).
\end{equation}
These counts also show that every rejected choice
meets a residue $1$ or $2$ at its first unsuccessful marker deletion.
The preceding values between deletions have residues $0$, $1$, or $2$.

On $b=(k-1)Q$, the first $k$ nonzero digits of $\tau_0$ occur
at exponents $Q,2Q,\ldots,kQ$, so each marker-one scan has length
$Q$. Set $H=1+p^Q+\cdots+p^{(k-1)Q}$. If $\tau_j$ is the
rescaled value after $j$ such scans, the recurrence
$\tau_j=\tau_{j-1}/p^Q-1$ gives
\[
 \tau_k=\frac{p^Q(H+1)}{1-p^Q-p^{kQ}}.
\]
In the coordinate $\tau=pz$, the map $T^{-1}$ acts as
$\tau\mapsto\tau/(1+\tau)$. Since
$1-p^Q-p^{kQ}+p^Q(H+1)=H$,
it sends $\tau_k$ to $p^Q(H+1)/H$, whose valuation is $Q$ and
leading residue is $2$.
The final scan therefore also has length $Q$. Its negated output
satisfies $px=2-(H+1)/H=(H-1)/H$. Using
$H=(1-p^{kQ})/(1-p^Q)$ gives
$x=p^{Q-1}(1-p^{(k-1)Q})/(1-p^{kQ})=X_{a,(k-1)Q-1}$.
Applying $T$ also gives $pT(x)=H-1=p^Q+\cdots+p^{(k-1)Q}$,
the case $m=k-1$ of the recovery identity \eqref{conf:finite-blocks} below.

We also need the initial digits remaining on the exceptional five-scan
branch. The rejection argument in \cref{contract:front-arithmetic}
uses these digits to determine the leading residue of its output.
Suppose $Q<b<2Q$ and put $d=b-Q$. The first five digits $1$
occur at exponents $Q,2Q,3Q,3Q+d,4Q$. The fifth marker is at
exponent $4Q$, so the five scans have total length $4Q$.
After these, the next contributions are
$2p^{4Q+d}$ and $p^{5Q}$, and all remaining terms have exponent
at least $5Q+d$. The same congruence, followed by deletion
of those five digits, therefore gives the value before $T^{-1}$ as
\begin{equation}\label{unary:common-front-false-prefix}
 \tau_5=2p^d+p^Q+u,
\end{equation}
where $\vp(u)\ge Q+d$.
\end{proof}

To recover a unary output from the intermediate code, we use
\begin{equation}\label{conf:finite-blocks}
 T\bigl(X_{a,m(a+1)-1}\bigr)
   =(1-p^{m(a+1)})z_a
   =\sum_{j=1}^{m}p^{j(a+1)-1},
\end{equation}
for $a\in\N$ and integers $m\ge1$. This follows by substitution
in \eqref{conf:two-field} and summing the finite geometric series.
The result encodes $m$ finite copies of $0^a1$, and a marker-one
scan removes one copy. For division, $m=k-1$. Removing $k-2$
copies leaves $p^a$, which $T$ sends to $z_a$.

\subsection{Division by two}

Apply \cref{unary:common-front} with $k=2$. Its accepted output
is $x=X_{a,a}$, so \eqref{conf:finite-blocks} gives
$T(x)=p^a$, which encodes the single finite block $0^a1$. A second $T$
returns $z_a$. Write
$K=T^2\mathcal N=M_{-1,2}$, so $K(z)=-z/(1+2pz)$.
The comparison's final generator normally applies $\mathcal N$ after
the marker-two deletion to produce $x$. Replacing that map by
$K=T^2\mathcal N$ also applies $T^2$ to the resulting $x$, giving
$z_a$ within the same generator.

\begin{lemma}\label{unary:odd-halving}
A module implements precisely $\mathsf{Div}_2(2a+1)=a$, for
$a\in\N$, on the unary codes. It has ten listed generators.
\end{lemma}

\begin{proof}
For boundaries $q,r$, reserve private intermediate states $g,u,v,w$.
The state $g$ is used while choosing the prefix length, $u$ and $v$
for the two marker-one scans, and $w$ for the final marker-two scan.
Supply the following ten generators.
\begin{equation}\label{unary:half-table}
 \begin{array}{ll}
 {}[q,g;0;\mathrm{id},\mathrm{id}],&[q,u;0;\mathrm{id},T],\\
 {}[g,g;0;\mathrm{id},\mathrm{id}],&[g,u;0;\mathrm{id},T],\\
 {}[u,u;0;\mathrm{id},\mathrm{id}],&[u,v;1;\mathrm{id},\mathrm{id}],\\
 {}[v,v;0;\mathrm{id},\mathrm{id}],&[v,w;1;\mathrm{id},T^{-1}],\\
 {}[w,w;0;\mathrm{id},\mathrm{id}],&[w,r;2;\mathrm{id},K].
 \end{array}
\end{equation}
On input $z_n$, let $b$ be the chosen number of zero deletions.
The direct edge $q\to u$ chooses $b=1$, while longer choices pass
through $g$. In either case, the generator entering $u$ combines
the last zero deletion with $T$. The marker exits $u\to v$ and
$v\to w$ complete the two marker-one scans. The latter also applies
$T^{-1}$ before the marker-two scan at $w$. Its exit $w\to r$
combines the marker deletion and recovery map in one generator.

For $1\le b\le n$, set $a=n-b$. By \cref{unary:common-front},
the comparison completes while remaining in $V$ exactly when
$b=a+1$, equivalently $n=2a+1$. The recovery calculation above
then gives output $z_a$. For every such input, choosing $b=a+1$
gives a path completing the instruction while remaining in $V$.

At $n=0$, the first zero deletion fails. A prefix longer than $n$
meets residue $1$ and also fails. Lemma~\ref{unary:common-front}
rejects every other incorrect guess and gives local residues in
$\{0,1,2\}$ before the rejecting deletion. The correct scans have
the same residue bound until the final generator returns a unary code,
whose residue is zero or one. Thus the state and digit checks remain
valid even for incorrect guesses, preventing a word from switching
instructions to avoid their rejection.
\end{proof}

\subsection{Division by three}

A successful comparison with $k=3$ leaves two finite blocks after applying
$T$, so recovery needs one additional marker scan.

\begin{lemma}\label{unary:ternary-division}
A module implements precisely $\mathsf{Div}_3(3a+2)=a$, for
$a\in\N$, on the unary codes. It has fourteen listed generators.
\end{lemma}
\begin{proof}
For boundaries $q,r$, reserve private intermediate states
$g,u_1,u_2,u_3,w,h$. The state $g$ is used while choosing the prefix
length, the $u_i$ for the three marker-one scans, $w$ for the
marker-two scan, and $h$ for the additional recovery scan.
Supply the following generators.
\begin{equation}\label{unary:third-table}
 \begin{array}{ll}
 {}[q,g;0;\mathrm{id},\mathrm{id}],&[q,u_1;0;\mathrm{id},T],\\
 {}[g,g;0;\mathrm{id},\mathrm{id}],&[g,u_1;0;\mathrm{id},T],\\
 {}[u_1,u_1;0;\mathrm{id},\mathrm{id}],&[u_1,u_2;1;\mathrm{id},\mathrm{id}],\\
 {}[u_2,u_2;0;\mathrm{id},\mathrm{id}],&[u_2,u_3;1;\mathrm{id},\mathrm{id}],\\
 {}[u_3,u_3;0;\mathrm{id},\mathrm{id}],&[u_3,w;1;\mathrm{id},T^{-1}],\\
 {}[w,w;0;\mathrm{id},\mathrm{id}],&[w,h;2;\mathrm{id},T\mathcal N],\\
 {}[h,h;0;\mathrm{id},\mathrm{id}],&[h,r;1;\mathrm{id},T].
 \end{array}
\end{equation}
The entry edges choose a nonempty zero prefix as in division by two.
Let $b$ be its length and, for $1\le b\le n$, set $a=n-b$.
The marker exits through $u_1,u_2,u_3$ count three markers equal
to one. The last of these, $u_3\to w$, also applies $T^{-1}$
before the marker-two scan. The exit $w\to h$ then combines the
marker-two deletion with $T\mathcal N(z)=-z/(1+pz)$, a permitted
auxiliary map that applies $T$ after the comparison's final negation.
By \cref{unary:common-front}, a path remaining in $V$ first reaches
$h$ exactly when $b=2(a+1)$, hence $n=3a+2$, and arrives with
value $T(X_{a,2a+1})$.
At $n=0$ the first zero deletion fails, and a prefix longer than $n$
fails at the unary marker. The common comparison excludes all remaining
incorrect guesses before arrival at $h$, since $T(E)\subseteq E$.

By \eqref{conf:finite-blocks}, the value on first arrival at $h$
represents two finite copies of $0^a1$. The loop at $h$ deletes the
$a$ zeros of the first copy, and the exit $h\to r$ deletes its
marker and applies $T$ to the remaining block. Thus
$D_1D_0^aT(X_{a,2a+1})=p^a$, followed by $T(p^a)=z_a$.
The common comparison supplies the residue bounds before that arrival,
and $T$ preserves residues. During the final scan the local residues
are $0$ or $1$, so the state and digit checks remain valid throughout.
Choosing $b=2(a+1)$ gives a path completing the instruction while
remaining in $V$ for every $a\ge0$, and every such path has the stated
output.
\end{proof}

\section{Contraction instruction modules}\label{sec:contractions}

This appendix gives the matrix constructions for the contraction instructions in
\cref{conf:instruction-table}. We first reject incorrect splits and
remove finite blocks, then recover the required unary output.
We use the local values and generator-table conventions of
Appendix~\ref{sec:unary}, together with the rescaled value $\tau=pz$
and maps $T,T^{-1}$ from Appendix~\ref{unary:comparison-proof}.
The arithmetic calculations hold for every prime $p\ge7$.
When applying the state checks in \cref{conf:guard} to unrestricted
matrix words, we also require $p>6N$, where $N$ counts all program
and intermediate states. The tables describe the intended graph paths,
which these checks enforce. Stages of one instruction may share an
intermediate state at a specified join, but intermediate states of
different instruction modules remain distinct.

\subsection{Comparison and block removal}
\label{contract:fivefold}

For local input $z_n$ with $n\ge1$, choose a zero prefix of length
$1\le b\le n$, and put $a=n-b\ge0$ and $Q=a+1$.
The comparison in \cref{unary:common-front} with $k=5$ permits
both the intended choice $b=4Q$ and the incorrect choices $Q<b<2Q$.
We will reject the latter when removing the first finite block.
Let $\iota$ be the entry state and $h_1$ the private intermediate
state where block removal begins. Write $x$ for the comparison's
output after negation. Its last generator also applies $T$, so the
local value at $h_1$ is $T(x)$.

\begin{lemma}\label{contract:front-arithmetic}
For the chosen prefix length $b$, a path from $\iota$ with local
input $z_n$ to $h_1$ remaining in $V$ exists exactly when
$b=4Q$ or $Q<b<2Q$. Every such path has local output $T(x)$.
If $b=4Q$, then $x=X_{a,4a+3}$ and the input is $z_{5a+4}$.
If $Q<b<2Q$, then $x$ is nonzero and its leading residue
is $4$, $3$, or $-1$, according as $2b<3Q$, $2b=3Q$, or
$2b>3Q$.

Along every path remaining in $V$, the local input and
the local value after each generator have residues in
$\Delta=\{-1,0,1,2,3,4\}$.
This includes $T(x)$ and values arising from incorrect guesses
before rejection.
\end{lemma}

\begin{proof}
Lemma~\ref{unary:common-front} with $k=5$ gives exactly these two
cases for completing the comparison in $V$. On $b=4Q$, its output is
$x=X_{a,4Q-1}=X_{a,4a+3}$ and $n=a+4Q=5a+4$.

For the additional branch, put $d=b-Q$, so $0<d<Q$.
By \eqref{unary:common-front-false-prefix}, the rescaled local value
after the five marker-one scans is $\tau_5=2p^d+p^Q+u$, where
$\vp(u)\ge Q+d$. Applying $T^{-1}$ changes this rescaled value to
$\tau_5/(1+\tau_5)$, which has valuation $d$ and leading residue two.
The local value has valuation $d-1$, so the marker-two scan deletes
$d-1$ zeros and then the marker. Its length is therefore $d$, and
negation gives
$px=2-p^{-d}\tau_5/(1+\tau_5)$. Multiplying by $1+\tau_5$ yields
\begin{equation}\label{conf:false-candidate}
 \begin{aligned}
 px(1+\tau_5)
 &=2(1+\tau_5)-p^{-d}\tau_5\\
 &=4p^d+2p^Q-p^{Q-d}+2u-p^{-d}u\\
 &\equiv4p^d-p^{Q-d}\pmod{p^Q}.
 \end{aligned}
\end{equation}
Since $1+\tau_5$ has
residue one, the leading residue of $x$ is $4$ when $2d<Q$,
$-1$ when $2d>Q$, and $3$ when $2d=Q$. These are the cases
$2b<3Q$, $2b>3Q$, and $2b=3Q$, respectively.
In particular, $\vp(x)=\min\{d,Q-d\}-1\ge0$, so $x\in V$
and $x\ne0$. None of these leading residues is zero or one when $p\ge7$.

By \cref{unary:common-front}, the local residues before the final
marker-two deletion belong to $\{0,1,2\}$. An incorrect candidate
$x$ has residue zero if $\vp(x)>0$, and otherwise has its leading
residue $3$, $4$, or $-1$. The correct candidate $X_{a,4a+3}$ has
residue zero or one. Finally, $T(x)=x/(1-px)$ preserves the residue
of every $x\in V$, and the valuation and leading residue when
$x\ne0$. Thus the final generator also gives a local residue in $\Delta$.
\end{proof}

\smallskip
\noindent\textbf{Comparison generators.}
We now give the graph and its sixteen listed generators. Reserve
distinct private intermediate states $g,u_1,u_2,u_3,u_4,u_5,w$.
The state $g$ is used while choosing the prefix length, $u_i$ for
the $i$th marker-one scan, and $w$ for the marker-two scan.
The row indexed by $i$ supplies eight of the sixteen edges.
\begingroup
\renewcommand{\arraystretch}{1.15}
\begin{equation}\label{contract:front-table}
 \begin{array}{@{}lll@{}}
 \text{Zero deletion}&\text{Next stage}&\text{Range}\\
 \midrule
 {}[\iota,g;0;\mathrm{id},\mathrm{id}]&[\iota,u_1;0;\mathrm{id},T]&\\
 {}[g,g;0;\mathrm{id},\mathrm{id}]&[g,u_1;0;\mathrm{id},T]&\\
 {}[u_i,u_i;0;\mathrm{id},\mathrm{id}]&[u_i,u_{i+1};1;\mathrm{id},\mathrm{id}]&1\le i<5,\\
 {}[u_5,u_5;0;\mathrm{id},\mathrm{id}]&[u_5,w;1;\mathrm{id},T^{-1}]&\\
 {}[w,w;0;\mathrm{id},\mathrm{id}]&[w,h_1;2;\mathrm{id},T\mathcal N].&
 \end{array}
\end{equation}
\endgroup
The direct edge $\iota\to u_1$ chooses $b=1$, while longer guesses
pass through $g$. Each edge entering $u_1$ deletes the last guessed
zero and then applies $T$. At $n=0$, the first deletion enters $E$,
and any choice $b>n$ also enters $E$ when it tries to delete the
unary code's marker as a zero. The auxiliary map
$T\mathcal N(z)=-z/(1+pz)$ is $M_{-1,1}$ from
Section~\ref{conf:control-states} and is part of the last generator.

\smallskip
\noindent\textbf{Removing finite blocks.}
For the correct candidate, $T(x)$ encodes four finite copies of
$0^a1$, as in \eqref{conf:finite-blocks}. A marker-one scan removes
one copy and rejects an incorrect candidate when it reaches its first
nonzero digit. Reserve a private intermediate output state $\omega$.
The following two listed generators remove one copy and apply
$T^{-1}$ to the remaining finite blocks.
\begin{equation}\label{contract:single-subtraction-table}
 [h_1,h_1;0;\mathrm{id},\mathrm{id}]
 \quad\text{and}\quad
 [h_1,\omega;1;\mathrm{id},T^{-1}].
\end{equation}
Alternatively, reserve one further private intermediate state $h_2$ and supply
these four listed generators for two successive block-removal scans.
\begin{equation}\label{contract:double-subtraction-table}
 \begin{array}{ll}
 {}[h_1,h_1;0;\mathrm{id},\mathrm{id}]&[h_1,h_2;1;\mathrm{id},\mathrm{id}],\\
 {}[h_2,h_2;0;\mathrm{id},\mathrm{id}]&[h_2,\omega;1;\mathrm{id},T^{-1}].
 \end{array}
\end{equation}
The state $\omega$ is where unary recovery begins. Reaching it does
not yet complete the contraction instruction.

\begin{proposition}\label{contract:front-module}
Adjoining \eqref{contract:single-subtraction-table} to
\eqref{contract:front-table} gives eighteen listed generators for
$z_{5a+4}\mapsto X_{a,3a+2}$.
Adjoining \eqref{contract:double-subtraction-table} instead gives
twenty listed generators for $z_{5a+4}\mapsto X_{a,2a+1}$.
In either case, a path from $\iota$ with local input $z_n$ to $\omega$
remaining in $V$ exists exactly when $n=5a+4$ for some $a\in\N$,
and every such path has the stated local output.
\end{proposition}

\begin{proof}
A path in \eqref{contract:front-table} that reaches $h_1$ performs
a nonempty guessed prefix, five marker-one scans, and a marker-two
scan. If it remains in $V$, Lemma~\ref{contract:front-arithmetic}
classifies its local value $T(x)$. For an incorrect candidate this
value has leading residue $3$, $4$, or $-1$. Taking the marker-one
exit while the residue is still zero enters $E$. After deleting all
the initial zeros, both another zero deletion and the marker-one exit
enter $E$. Thus no incorrect candidate can reach $\omega$ while
remaining in $V$, and its local residues before rejection belong to
$\{0,3,4,-1\}$.

On the correct split, the local value at $h_1$ is $T(X_{a,4a+3})$.
By \eqref{conf:finite-blocks}, this value encodes the finite word consisting
of four copies of $0^a1$. A marker-one scan removes its first copy,
with exactly $a$ zero deletions. In the single-scan table, the final
$T^{-1}$ therefore recodes the three remaining copies as
$X_{a,3a+2}$. In the double-scan table, the first scan leaves
those three copies at $h_2$. The second scan removes one
more copy, and its final $T^{-1}$ gives $X_{a,2a+1}$.
Neither branch deletes the last copy, so all these values remain
nonzero. Every correct scan has local residues in $\{0,1\}$, including
$a=0$, when each block consists only of its marker.

These residue bounds also let us apply \cref{conf:guard}. Once a
word enters this instruction's private intermediate states, it
cannot use a generator for another state to skip the remaining
tests while staying in $V$. Conversely, choosing $b=4(a+1)$
reaches the asserted output for every $a\ge0$.
\end{proof}

\subsection{Recovering the unary output}
\label{contract:projection-two}

The preceding block-removal construction reaches recovery with
$X_{a,m(a+1)-1}$, where $m\in\{2,3\}$. We now recover the unary
code $z_{m(a+1)-1}$, as described in Section~\ref{conf:output-recovery},
and prove that the final scans reject every incorrect recovery guess.

Put $Q=a+1$ and $S_m=1+p^Q+\cdots+p^{mQ}$. A marker-one scan
brings the longer zero block to the front. It has length $Q$ and gives
\[
 Y=D_1D_0^a(X_{a,mQ-1})=-X_{mQ-1,a}
 \quad\text{and}\quad
 pY=-\frac{p^{mQ}}{S_m}.
\]
We aim to subtract $p^Q,\ldots,p^{(m-1)Q}$ from $S_m$, leaving
$1+p^{mQ}$. A guessed zero prefix followed by $T^{-1}$ subtracts
one power of $p$, so we make $m-1$ such guesses and then check them.

Put $\tau_0=pY$ and $r_0=mQ$. As long as the local values remain
in $V$, write $\tau_i$ for the rescaled value after the $i$th guessed
prefix and the following application of $T^{-1}$, and put
$r_i=\vp(\tau_i)$. We show that
\begin{equation}\label{conf:projection-guesses}
 \tau_i=-\frac{p^{r_i}}{S_m-\sum_{j=1}^{i}p^{r_j}},
 \qquad 1\le i<m,
\end{equation}
and that the choices whose deletions remain in $V$ are exactly
$1\le r_{m-1}<\cdots<r_1<r_0=mQ$. Here $r_i$ is the remaining
exponent in the rescaled value $\tau_i$. The length of the $i$th
guessed prefix is $r_{i-1}-r_i$.

Initially, $S_m$ has residue one. After any guesses with positive
exponents, the denominator in \eqref{conf:projection-guesses} still
has residue one. Before the $i$th guess, the local value therefore
has $r_{i-1}-1$ leading zeros and leading residue $-1$.
A nonempty zero prefix can lower the exponent to any $r_i$ with
$1\le r_i<r_{i-1}$. The following $T^{-1}$ preserves this valuation
and leading residue. On the rescaled value $\tau=pz$, it acts as
$\tau\mapsto\tau/(1+\tau)$, adding the negative numerator to the
denominator. Thus the two operations give
\[
 \begin{aligned}
 -\frac{p^{r_{i-1}}}{S_m-\sum_{j=1}^{i-1}p^{r_j}}
 &\xrightarrow{\text{delete }r_{i-1}-r_i\text{ zeros}}
 -\frac{p^{r_i}}{S_m-\sum_{j=1}^{i-1}p^{r_j}}\\
 &\xrightarrow{T^{-1}}
 -\frac{p^{r_i}}{S_m-\sum_{j=1}^{i}p^{r_j}}.
 \end{aligned}
\]
Induction proves \eqref{conf:projection-guesses} and the stated
exponent bounds. Conversely, every sequence within those bounds is obtained by deleting
$r_{i-1}-r_i$ zeros at its $i$th guess.

After the last guess, collect the remaining nonconstant terms of the
denominator in
\begin{equation}\label{conf:projection-candidate}
 H=\sum_{j=1}^{m}p^{jQ}-\sum_{i=1}^{m-1}p^{r_i}.
\end{equation}
The denominator is now $1+H$. The scan to the marker $-1$ has
length $r_{m-1}$ and sends the rescaled value to
$-1/(1+H)+1=H/(1+H)$.
The first application of $T$ turns this rescaled value into
$H$, and the second gives $\tau=H/(1-H)$. Their composition is
the permitted map $T^2(z)=z/(1-2pz)=M_{1,-2}(z)$.
We apply $m$ successive marker-one scans to the candidate local
value $z=\tau/p$.

\begin{lemma}\label{contract:projection-arithmetic}
Let $m\in\{2,3\}$, $Q\ge1$, and
$1\le r_{m-1}<\cdots<r_1<mQ$ be integers. For the candidate local
value $z=\tau/p$, where $\tau=H/(1-H)$ and $H$ is defined in
\eqref{conf:projection-candidate}, all $m$ marker-one scans
finish without entering $E$ exactly when $r_i=(m-i)Q$ for $1\le i<m$, or equivalently
when each guessed prefix has length $Q$. The accepted local value is
$z_{mQ-1}$, and each scan preserves it.

Throughout the recovery
procedure, the local values at the input and after every generator
application along a path remaining in $V$, including those reached
by incorrect guesses before rejection, have residues in
$\{-1,0,1,2\}$.
\end{lemma}

\begin{proof}
If every guessed exponent is divisible by $Q$, their allowed range
and strict ordering force
$r_i=(m-i)Q$. Then $H=p^{mQ}$ and
$\tau=p^{mQ}/(1-p^{mQ})=p z_{mQ-1}$, so the candidate local
value is $z_{mQ-1}$ and every marker-one scan preserves it.

\smallskip
\noindent\textbf{Rejecting incorrect recovery guesses.}
For an incorrect guess, cancel matching positive and negative powers
in $H$, and let $-p^D$ be the remaining negative term of least
exponent. Such a term exists because the allowed range contains exactly
$m-1$ positive multiples of $Q$, and the guesses are distinct.
In particular, $Q\nmid D$ and $D<mQ\le3Q$.

Since $H\in p\Z$, the finite geometric identity gives
\begin{equation}\label{contract:projection-initial-digits}
 \tau=\sum_{j=1}^D H^j+\frac{H^{D+1}}{1-H}
   \equiv\sum_{j=1}^D H^j\pmod{p^{D+1}}.
\end{equation}
Below $D$, only the surviving terms
$p^Q$, $p^{2Q}$, and the product of two $p^Q$ terms can contribute.
Among exponents below $D$, the coefficients at $Q$ and $2Q$ are at most
one and two, respectively. Products of positive terms have exponents
divisible by $Q$, so they cannot contribute at $D$, since $Q\nmid D$.
Any other product containing a negative term has larger exponent.
The coefficient at $D$ is therefore $-1$. With $p\ge7$ there are
no carries before $D$.

Passing from $\tau$ to the local value $z=\tau/p$ lowers all
exponents by one without changing these coefficients. Each successful
marker-one scan removes the next zeros and coefficient one.
At a coefficient two or minus one, both another zero deletion
and the marker-one exit enter $E$. There are at most $m-1$ positive
multiples of $Q$ below $D$, so at most $m-1$ coefficients one can
precede rejection. The $m$ marker-one scans therefore reject every incorrect
candidate.

\smallskip
\noindent\textbf{Residues during recovery.}
During the initial scan of $X_{a,mQ-1}$, the local residues are zero or one.
The guessed prefixes and following scan to the marker $-1$ have only zero
and minus one before that marker exit, by \eqref{conf:projection-guesses}.
The coefficient argument gives only local residues zero, one, two, or minus one during
the final $m$ marker-one scans.
The nonzero coefficient at $D$ prevents an incorrect candidate from
becoming zero before rejection. Thus all required local residues are
in the stated set. When $a=0$, we have $Q=1$, and the inequalities
$1\le r_{m-1}<\cdots<r_1<m$ force $r_i=m-i$.
\end{proof}

\smallskip
\noindent\textbf{Recovery generators.}
Let $\iota$ be the intermediate state joining the block-removal and
recovery stages of this instruction. Every local value entering from
block removal along a path remaining in $V$ has the form $X_{a,mQ-1}$.
Let $\omega\ne\iota$ be the exit state, where the unary output is produced.
The initial scan starts at $\iota$, which has only that scan's
outgoing edges. Reserve $2m$ further private intermediate states
$g_1,\ldots,g_{m-1},\ell,v_1,\ldots,v_m$.
For the table only, write $g_m=\ell$ and $v_{m+1}=\omega$.
Supply exactly the following $4m+2$ edges.
\begingroup
\renewcommand{\arraystretch}{1.15}
\begin{equation}\label{contract:projection-table}
 \begin{array}{@{}lll@{}}
 \text{Zero deletion}&\text{Next stage}&\text{Range}\\
 \midrule
 {}[\iota,\iota;0;\mathrm{id},\mathrm{id}]&[\iota,g_1;1;\mathrm{id},\mathrm{id}]&\\
 {}[g_i,g_i;0;\mathrm{id},\mathrm{id}]&[g_i,g_{i+1};0;\mathrm{id},T^{-1}]&1\le i<m,\\
 {}[\ell,\ell;0;\mathrm{id},\mathrm{id}]&[\ell,v_1;-1;\mathrm{id},T^2]&\\
 {}[v_i,v_i;0;\mathrm{id},\mathrm{id}]&[v_i,v_{i+1};1;\mathrm{id},\mathrm{id}]&1\le i\le m.
 \end{array}
\end{equation}
\endgroup
Each guessing exit deletes the last zero of its prefix before
applying $T^{-1}$. Hence every guess is nonempty, even if its
loop is unused, and every positive prefix length can be attempted.

\begin{proposition}\label{contract:projection-module}
For $m\in\{2,3\}$ and $a\in\N$, the construction
\eqref{contract:projection-table} has a path from $\iota$ to $\omega$
with local input $X_{a,m(a+1)-1}$ that remains in $V$.
Every such path gives local output $z_{m(a+1)-1}$.
After the global choice $p>6N$, the same conclusion holds for
unrestricted matrix words, with completion at their first arrival
at $\omega$. The construction has $4m+2$ listed generators.
\end{proposition}

\begin{proof}
An entry-to-exit path in the listed graph performs the first scan,
the $m-1$ nonempty guesses, the scan to the marker $-1$, and the
$m$ marker-one scans in that order.
Lemma~\ref{contract:projection-arithmetic} shows that a path remaining
in $V$ reaches the output only when each guessed prefix has length $Q$,
and gives the residue bounds, including those of incorrect candidates.
Choosing $r_i=(m-i)Q$ supplies a path reaching
$\omega$ while remaining in $V$ for every $a\ge0$, and the lemma gives the asserted
output for every such path. Once $p>6N$, these residue bounds let
\cref{conf:guard} enforce the state and digit checks along an
unrestricted matrix word. They cover incorrect guesses as well, so
a word remaining in $V$ cannot leave this construction to bypass
the remaining tests. Any such word reaching $\omega$ must therefore
follow the listed stages until its first arrival there.
\end{proof}

\subsection{Assembly and verification}
\label{contract:assembly}

\begin{corollary}\label{contract:contractions}
The updates $\mathsf{Con}_2$ and $\mathsf{Con}_3$ in
\cref{conf:instruction-table} have modules satisfying the three requirements
of Section~\ref{conf:module-conditions} for every prime $p\ge7$, with respectively thirty and
thirty-two listed generators.
\end{corollary}

\begin{proof}
Join the twenty-generator construction of
\cref{contract:front-module} to recovery with $m=2$, and its
eighteen-generator construction to recovery with $m=3$.
For each $m\in\{2,3\}$ and $a\in\N$, the two stages give
$z_{5a+4}\longmapsto X_{a,m(a+1)-1}\longmapsto z_{m(a+1)-1}$.
Thus the first stage produces exactly the intermediate code required
by recovery, and the final unary code records the output of
$\mathsf{Con}_m$.

Identify the private output state of the first stage with the entry
state of recovery, retaining only the recovery scan's outgoing edges
there and keeping all other intermediate states distinct.
By \cref{contract:front-module}, a path remaining in $V$ can first
reach this shared state only with the displayed intermediate code.
Subsequent zero deletions change that
form, but their arithmetic and residue bounds are covered by the
recovery proof. The shared state remains private, so reaching it
does not complete the instruction.
\Cref{contract:projection-module} then gives the exact unary output
and a path completing the instruction while remaining in $V$ for every
input in the domain.

The residue bounds and the guarantees for inputs and outputs verify
the corresponding requirements in Section~\ref{conf:module-conditions}.
The finite tables of permitted
generators can be constructed independently of $n$ and before choosing
$p$, which verifies the effective construction requirement.
The numbers of listed generators are $20+10=30$ and $18+14=32$.
\end{proof}

\begin{proof}[Proof of \cref{conf:instruction-implementations}]
Fix a prime $p\ge7$. For each instruction occurrence $q\xrightarrow{f}r$,
take a separate copy of its module from Appendix~\ref{sec:unary} or
\cref{contract:contractions}. Label its entry and exit by $q$ and $r$,
and give all other states distinct private labels. Only $q$ and $r$
are program boundaries. Completion means reaching the next boundary
after a positive number of edges, including when $q=r$.
We verify the three requirements of Section~\ref{conf:module-conditions}
for graph paths starting at the entry state with local input $z_n$,
for every $n\in\N$.

\paragraph{Inputs and outputs.}
For increment, congruence tests, and division,
\cref{unary:basic-instructions,unary:congruence-tests,unary:odd-halving,unary:ternary-division}
give exactly the domains and outputs in \cref{conf:instruction-table}.
For contractions, \cref{contract:contractions} gives the assembled
modules with the private joins specified in its proof. In each case,
the cited proofs give a path completing the instruction while
remaining in $V$ for every allowed input, and show that every such
completed path has an allowed input and the prescribed unary output.

\paragraph{Residues along every prefix.}
The residue requirement concerns every prefix remaining in $V$,
even after an incorrect guess or when the path never completes.
For increment and congruence tests, all local residues lie in
$\{0,1\}$ by \cref{unary:basic-instructions,unary:congruence-tests}.
For division, \cref{unary:common-front} gives residues in
$\{0,1,2\}$ before the comparison's final exit. After that exit,
the composed maps give either the unary output or finite blocks
whose remaining scan has residues zero and one, as proved in
\cref{unary:odd-halving,unary:ternary-division}.
For contractions, \cref{contract:front-arithmetic,contract:front-module}
place the comparison and block-removal residues in $\Delta$,
including those of incorrect candidates before rejection.
The recovery residues lie in $\{-1,0,1,2\}$ by
\cref{contract:projection-arithmetic}. These are all subsets of $\Delta$.
The bounds concern local values after complete generator applications.
They also apply across the private joins, since the preceding stage
supplies the input family required by the next stage, as established
in the proof of \cref{contract:contractions}.

\paragraph{Effective construction.}
Each cited construction supplies a finite graph and a list of
edge formulas of the form \eqref{conf:generator}. The parameters
$d,A$ of a test and the scan counts $k,m$ are fixed by the instruction.
Different counter values use different paths in these fixed graphs,
including different numbers of loop repetitions. Neither the graph
nor its formulas depend on $n$. The auxiliary maps and
their compositions have the permitted form, as verified in the
constructions, and the formulas give computable matrix entries once
$p$ is chosen. All the local calculations and residue bounds hold
for every prime $p\ge7$ with the same graphs and formulas.
This verifies all three requirements and proves the proposition.
\end{proof}

\section{Generator count and padding to an exact family size}
\label{sec:counts}

We prove \cref{conf:count} by first counting the listed generators.
If some entries represent the same matrix, we add extra matrices
to reach the required family size. This padding preserves reachability
between encoded configurations and is independent of the machine input $\rho$.

\subsection{Counting the listed generators}

The instruction counts of $U$ are given in \cref{prog:source-counts}.
Translating these instructions into $F$ and reversing its edges gives
the instructions of $P$. An increment of counter zero contributes
one $\mathsf{Div}_2$ instruction, and an increment of counter one
contributes one $\mathsf{Div}_3$ instruction. Each conditional decrement
contributes two instructions of $P$. Its positive branch gives
$\mathsf{Con}_2$ or $\mathsf{Con}_3$, according to the selected counter,
and its zero branch gives a congruence test. For counters zero and one,
the tests have parameters $(d,A)=(2,\{0\})$ and $(3,\{0,1\})$,
respectively. By \cref{unary:congruence-tests}, each test module has
$1+d+|A|+\mathbf1_{0\in A}$ listed generators, giving five and seven.
The cleanup edge reverses to one $\mathsf{Inc}$ instruction of $P$,
implemented by four listed generators (\cref{unary:basic-instructions}).
The division and contraction counts in \cref{tab:generator-count}
come from \cref{unary:odd-halving,unary:ternary-division,contract:contractions}.

\begin{table}[htbp]
\centering
\begingroup
\normalsize
\setlength{\belowcaptionskip}{6pt}
\caption{Numbers of listed generators. Distinct table entries may
represent the same matrix.}
\label{tab:generator-count}
\renewcommand{\arraystretch}{1.1}
\setlength{\tabcolsep}{6pt}
\begin{tabularx}{\linewidth}{@{}>{\raggedright\arraybackslash}Xrrr@{}}
\toprule
\textbf{Instruction or construction step} & \textbf{Number}
 & \textbf{\shortstack[r]{Listed generators\\per occurrence}} & \textbf{Total}\\
\midrule
Increment counter 0 & 88 & 10 & 880\\
\addlinespace[3pt]
Increment counter 1 & 73 & 14 & 1,022\\
\addlinespace[3pt]
Conditional decrement counter 0 & 80 & $30+5$ & 2,800\\
\addlinespace[3pt]
Conditional decrement counter 1 & 28 & $32+7$ & 1,092\\
\addlinespace[3pt]
Reversed cleanup & 1 & 4 & 4\\
\midrule
Listed nonsingular generators & & & 5,798\\
\addlinespace[3pt]
Rank-one conversion & 1 & 1 & 1\\
\midrule
Total listed generators & & & 5,799\\
\bottomrule
\end{tabularx}
\endgroup
\end{table}

\subsection{Padding to exactly 5,798 distinct matrices}

\begin{lemma}[Exact cardinality]\label{counts:alphabet}
The fixed program $P$ in \cref{prog:universal} can be implemented by
exactly 5,798 distinct nonsingular integer matrices of determinant
$p$ or $-p$, for one fixed prime $p$, without changing reachability
between its encoded configurations.
\end{lemma}
\begin{proof}
Let $N$ count the program and intermediate states in the original
instruction graphs, indexed by $0,\ldots,N-1$. Reserve unused state
indices $N,\ldots,N+5797$, then choose a prime $p>6(N+5798)$.
This separates all residue blocks as required by
Section~\ref{conf:control-states} and leaves room for any needed
additions before we know how many original matrices are distinct.
With $p$ fixed, compute the original generators and remove duplicates.
Let $g\le5798$ be the number of distinct matrices. On $5798-g$ of
the reserved state indices $d$, add
\begin{equation}\label{counts:dummy}
 B_d=[d,d;0;\mathrm{id},\mathrm{id}]
     =C_dD_0C_d^{-1}
     =\begin{pmatrix}1&(p-1)(6d+1)\\0&p\end{pmatrix}.
\end{equation}

These matrices have determinant $p$ and are pairwise distinct because
their upper-right entries differ. To distinguish $B_d$ from every
original generator, consider $C_d(0)$. Since $0\in V$ has residue zero,
every original generator sends this point into $E$ by \cref{conf:guard},
because it is intended for a state other than $d$.
In contrast, $B_d(C_d(0))=C_d(0)\in V$, so $B_d$ cannot equal an
original generator. The same lemma shows that each $B_d$ maps $E$
into itself.

We now show that adding these matrices creates no new reachability
between encoded configurations. Consider a word that starts from an
encoded program configuration and contains a padding matrix.
Let $B_d$ be its first padding matrix.
The preceding prefix uses only original generators. If it has entered
$E$, it stays there. Otherwise, the induction in
Appendix~\ref{unary:compiler-proof} shows that this prefix follows the
original instruction graphs. The residue requirement of
Section~\ref{conf:module-conditions} keeps each local residue in $\Delta$,
and the input and output requirement gives a unary code
at each program boundary, allowing the induction to continue.
Thus the current point is $C_q(z)$ for an original state $q$ and
$z\in V$ with residue in $\Delta$, even if the prefix ends halfway
through an instruction. Since $d\ne q$, the state check sends this
point into $E$ when $B_d$ is applied. Every subsequent original or
padding matrix keeps it there, so the word cannot end at an encoded
configuration. Conversely, every original word remains available.
Reachability is therefore unchanged, and the enlarged family has
exactly 5,798 distinct matrices. All choices depend only on the fixed
program $P$, not on the machine input $\rho$.
\end{proof}

\section{Related decidability and membership results}
\label{app:scope}
\label{complete:membership-scope}

For unrestricted mortality, rational and integer entries give equivalent
decision problems. Clearing each generator's denominators multiplies
every product by a nonzero scalar, preserving zero products without
adding generators. For two-by-two integer matrices, mortality is
NP-hard when determinants lie in $\{0,1,-1\}$
\citep{BellHirvensaloPotapov2012} and NP-complete when they lie in
$\{0,1\}$ \citep[Theorem~4]{JaaxKiefer2020}.
For comparison with our generator count, undecidability is known for
six three-by-three integer matrices \citep[Corollary~12]{Neary2015}.

Prescribing a target matrix gives a different question from prescribing
a target line. For a finite family $\mathcal A$ of nonsingular
two-by-two integer matrices and a given nonsingular integer matrix
$M$, it is decidable whether $M\in\mathcal A^*$
\citep[Theorem~2.1]{PotapovSemukhin2017Nonsingular}.
For our family $\mathcal G$, a product of length $\ell$ has determinant
of absolute value $p^\ell$, so equality with $M$ requires
$p^\ell=|\det M|$. This leaves at most one possible length and
finitely many words to check. Requiring $W(s)=t$, as in
\cref{conf:reachability}, instead specifies only the image of one
line. It does not determine $\det W$, so no such length bound follows.

Mortality has also been studied for finite automata whose transitions
carry rational two-by-two matrix labels. Here the automaton restricts
which sequences of matrices may be multiplied. An algorithm can decide
whether the automaton accepts a zero product, provided every transition
on a directed cycle is labelled by a product of integer matrices of
determinant $\pm1$,
rational scalar matrices, and
$s_0=\left(\begin{smallmatrix}1&0\\0&0\end{smallmatrix}\right)$
\citep[Definition~6.2 and Theorem~8.1]{DiekertPotapovSemukhin2024}.

All words over $\mathcal G\cup\{R_\rho\}$ can be represented by an
automaton with one initial and accepting state and a loop for each
generator. Every loop is a directed cycle, so every generator would
have to satisfy the restriction above. A nonsingular product of the
permitted matrices cannot contain a singular factor. It is therefore
a nonzero rational scalar multiple of an integer matrix of determinant
$\pm1$, so its determinant has absolute value equal to a rational
square. Each generator in $\mathcal G$ instead has determinant of
absolute value $p$, which is not a rational square. The mortality
theorem therefore does not apply directly to this automaton. This
argument does not exclude other presentations satisfying its hypothesis.

\section{Formal verification}
\label{app:formal-verification}

A companion Lean~4 formalization is available at
\href{https://github.com/j49liu/matrix_mortality_lean}
{github.com/j49liu/matrix\_mortality\_lean}. It constructs the concrete
family used in the proof: 5,798 distinct nonsingular integer matrices,
each with determinant $51{,}859$ or $-51{,}859$, and one rank-one
integer matrix computed from the input. Lean proves that the rank-one
matrix always has the same image line, that the resulting family has
exactly 5,799 distinct matrices, and that it is mortal exactly when a
fixed universal machine halts on the corresponding input. It also states
the final undecidability result with the rank-one matrix itself as the
input, matching \cref{conf:main}.

The main danger in the reduction is that the matrices may be
multiplied in any order, rather than in the order of a legal computation.
The formalization therefore proves the converse for every matrix word:
no wrong control state, digit, scan length, or arithmetic guess can carry
one encoded configuration to another. It also proves the rank-one
conversion for an arbitrary zero product, without assuming that the
rank-one matrix occurs only twice or in prescribed positions. These
checks rule out unintended products that could otherwise create a false
mortality witness.

To obtain unconditional undecidability, the development starts from
a kernel-checked universal machine, proves that the complete map from
machine inputs to matrix families is primitive recursive, and derives
the required prime system from formal Mertens and Chebyshev estimates.
Formalization made two previously implicit obligations explicit: the
connection to a proved undecidable halting problem and the existence of
the number-theoretic prime layout. Both are now proved in Lean. Lean
checks every theorem in the development; none is left unproved or
assumed as a project-specific axiom.The paper's mathematical arguments 
remain self-contained and do not rely on any formalized results that are 
not stated in the paper.

\section{Conclusion and open problems}
\label{sec:conclusion}

We have proved that matrix mortality is undecidable for $2\times2$
integer matrices, even when all but one matrix are fixed in advance.
The fixed nonsingular matrices implement the same program for every
input, while the input changes only the kernel of a rank-one matrix
whose image stays fixed. Thus, deciding whether a product from a fixed
matrix family sends one fixed line to a given target line is already
undecidable. The construction enforces the program through the matrix
actions themselves, without restricting the order in which the
matrices may be multiplied or how often they may be used.

The distinction between reaching a target line and producing a specified
matrix is important here. Requiring a product to send one line to
another leaves its action on the remaining directions unspecified.
By comparison, deciding whether a given nonsingular matrix is a
product of a finite family of nonsingular $2\times2$ integer matrices
is decidable. Appendix~\ref{app:scope} explains this distinction and
its relation to existing decidability results. Our theorem establishes
undecidability for one fixed family of nonsingular matrices. It does
not classify which fixed families admit an algorithm for deciding
reachability between rational lines.

The construction gives an explicit bound of 5,799 distinct matrices,
whereas mortality is decidable for families of two rational $2\times2$
matrices \citep[Theorem~4]{BournezBranicky2002}. Determining how many
matrices are needed for undecidability is therefore a natural next
question. There are two versions of this question that should be
distinguished. One allows every matrix in the family to vary with the
input. The other retains the restrictions proved here, with a fixed
nonsingular family and only the kernel of one rank-one matrix varying.
Our bound applies to this second, more restricted setting, but we do
not determine the smallest family size in either setting or whether
those sizes coincide.

A further question concerns the determinants of the nonsingular
matrices. Our construction works for every sufficiently large prime
$p$, with the required size depending on the number of states that
must receive separate residue blocks. Can mortality remain undecidable
when these determinants are restricted to $\{2,-2\}$, while the
nonsingular family and the image of the rank-one matrix remain fixed?
The present construction does not answer this question, since its
state checks require more residue classes than are available modulo
two. A different way to distinguish states might also allow fewer
intermediate states and fewer generators. Any such simplification
would still need to prevent a matrix word from switching instructions
to avoid a test, including after an incorrect arithmetic guess.

\section{AI methodology}
\label{app:ai-methodology}
We used OpenAI Codex and Claude Code for mathematical research, proof development, manuscript preparation, and Lean~4 formalization. The investigation began in July 2026, when the authors used GPT-5.6-Sol to systematically investigate this problem. This earlier investigation produced a large archive of partial results, but did not resolve the problem. The final proof architecture was selected by the authors and developed iteratively with AI assistance. OpenAI Codex with GPT-5.6-Sol was also used to generate a Lean formalization of the proofs. The authors substantially reworked the exposition into its current form, reviewed the correspondence between the formalization and the mathematical statements, checked the arguments and references, and revised the manuscript. The authors take full responsibility for the correctness, originality, and integrity of the submission.

\bibliographystyle{alpha}
\bibliography{references}
\end{document}